\documentclass[twocolumn,showpacs,preprintnumbers,amsmath,amssymb,aps]{revtex4}
\usepackage{graphicx}
\begin{document}

\title{Switching and Jamming Transistor Effect for Vortex Matter
in Honeycomb Pinning 
Arrays with ac Drives}  
\author{C. Reichhardt  
and C.J. Olson Reichhardt} 
\affiliation{ 
Theoretical Division, 
Los Alamos National Laboratory, Los Alamos, New Mexico 87545}
\date{\today}
\begin{abstract}

We show that a remarkable variety of dynamical phenomena, including 
switching, polarization, symmetry locking, and dynamically induced 
pinning, can occur for 
vortices in type-II superconductors in the presence of 
a honeycomb pinning array and an ac or combined ac and dc drive.
These effects
occur at the second matching field where there are two vortices per pinning
site, and arise due to the formation of  
vortex dimer states in the interstitial regions of the honeycomb array.   
The orientation of the pinned and moving vortex dimers can be 
controlled externally by 
the application of a drive.  We term this a polarization effect and demonstrate
that it can lock or unlock the vortex motion into different  
symmetry directions of the underlying pinning lattice.   
If the moving vortices are locked into one direction, the motion can be
switched into a different direction by applying an additional bias drive,
producing sharp jumps in the transverse and longitudinal 
velocities. 
Further, the dc vortex motion in one direction can be controlled directly 
by application of a force in the perpendicular direction. 
When the moving dimers reorient, we find
a remarkable dynamical pinning effect 
in which the dimers jam when they
become perpendicular to the
easy flow direction of the pinning lattice. 
Since application of an external field can be used to 
switch off the vortex flow, we term
this a jamming transistor effect.  
These effects do not occur in triangular pinning arrays due to the lack of the 
$n$-merization of the vortices in this case.
The switching and dynamical pinning effects demonstrated here
may be useful for the creation of new types of fluxtronic devices.  
\end{abstract}
\pacs{74.25.Qt}
\maketitle

\section{Introduction}

The dynamics of vortices in type-II superconductors has attracted enormous 
interest since they can provide insight into 
nonequilibrium phases of driven particles moving over 
a quenched landscape 
and
properties of vortex  dynamics
that can be useful
for creating new types of fluxtronic devices. 
In terms of basic science, driven vortices 
are an ideal system in which to study
different types of nonequilibrium phase transitions 
\cite{Brandt,Koshelev,Giamarchi,Balents,Moon,Higgins,Pardo}. 
It appears that 
nonequilibrium systems can undergo transitions between
specific types of driven phases similar to the 
transitions among
equilibrium phases that occur as a parameter is tuned. 
Vortices driven over a randomly 
disordered substrate can pass through
a pinned phase, a disordered or plastic flow phase with large fluctuations
\cite{Brandt,Moon,Higgins}, 
and different types of moving phases
which may be partially ordered in one 
direction to form a moving smectic \cite{Balents,Moon,Pardo} 
or have moving crystal symmetry \cite{Giamarchi,Balents,Pardo}. 
Other types of phases are also possible within the plastic flow 
regime, including chaotic \cite{Soret},
reversible, and irreversible flows \cite{Mangan}. 
For driven vortices in samples with periodic pinning arrays, 
an exceptionally rich variety of
distinct dynamical behaviors appear
\cite{Reichhardt,Zimanyi,Carneiro,Zhu,Harada,Rosseel,Jensen1,Velez,Sur,Nori,Velez1,Sil,Velez2,Wun,Jiang,Nishio}. 
Periodic pinning arrays in superconductors can be created with
nanoholes 
\cite{Fiory,Metlushko,Welp,Bending,Field,Commensurate,Peeters1,Karapetrov,Peeters,Pannetier,Jensen,Horng,Wu} 
or magnetic dots \cite{Martin,Fertig,Morgan}. 
Typically, strong matching effects are observed in these samples
as anomalies in the critical current or other transport 
measurements when the number of vortices equals an integer multiple or 
rational fraction of the number of pinning sites.    
At the matching field $B_\phi$, there are an equal number of vortices
and pins.
Above the first matching field, 
the additional vortices corresponding to the excess density $B-B_\phi$
can be located in the interstitial regions between the occupied pinning
sites already occupied by vortices \cite{Harada,Bending,Field,Reichhardt,Karapetrov,Peeters} 
or multiple vortices can be trapped at each pinning 
site \cite{Bending,Field,Peeters,Pannetier,Jensen}. 
If interstitial vortices are present, they are more mobile than the pinned
vortices under an applied drive and have 
a lower depinning threshold, so the interstitial vortices 
can move around the pinned vortices as
observed in experiments \cite{Harada,Rosseel} and simulations 
\cite{Reichhardt,Zhu}.  
The interstitial vortex motion can be probed 
by adding an ac drive to a dc drive to produce Shapiro steps
in the current-voltage curves \cite{Rosseel}.  
The symmetry of the underlying pinning array can lock the vortex motion
into particular directions
when the vortices are driven at various angles with respect to the
pinning lattice, as predicted in simulations \cite{Nori} 
and demonstrated in experiment \cite{Velez1,Sil,Velez2}. 
In addition to the superconducting system, vortices interacting with  
periodic substrates can also be realized in  
Bose-Einstein condensates with a 
periodic optical trap array \cite{Bigelow,Demler,Tung}, 
opening a new avenue for the study of vortex matter on dynamical substrates.  

Studies of vortices interacting with periodic pinning arrays have primarily 
concentrated on 
triangular or square arrays; however, 
other types of pinning geometries such as 
honeycomb or kagom{\' e} arrays 
can also be realized in experiment \cite{Wu,Morgan}. 
A honeycomb pinning array is constructed 
from a triangular array by removing $1/3$ 
of the pinning sites, while
a kagom{\' e} array is generated by removing $1/4$ of the pinning sites.
Experiments have shown that honeycomb and kagome pinning arrays exhibit 
anomalies in the critical currents at 
non-matching fields as well as at matching fields \cite{Wu,Morgan}. 
In honeycomb pinning arrays, non-matching anomalies occur at 
fields $B/B_\phi=n + 1/2$, where 
$n$ is an
integer. This suggests that a portion of the vortices 
are trapped in the large interstitial regions in such 
a way as to match the original undiluted triangular pinning array. 
Simulations also reveal matching effects 
at integer and half integer matching 
fields up to $B/B_\phi=5$  for 
honeycomb arrays, 
and 
indicate that this type of matching can occur when
multiple interstitial vortices are trapped at the large interstitial 
sites in the honeycomb array and form either ordered vortex molecular crystals
or vortex plastic crystals \cite{Pinning}.
In vortex molecular crystals, the vortices in the large 
interstitial site behave as $n$-mer objects that have a director field 
which can be oriented in different directions with respect to 
the pinning lattice.
The $n$-mers can have long range order and may experience
ferromagnetic coupling that aligns the $n$-mers in 
a single direction, antiferromagnetic coupling that causes neighboring
$n$-mers to orient perpendicularly to each other, 
or other types of coupling that produce ordered phases \cite{Pinning}.       
Vortex molecular crystals form at integer and half integer 
matching fields when $B/B_{\phi} \geq 2$ for honeycomb pinning arrays 
and at integer multiples of $B/B_\phi=1/3$ for kagome 
pinning arrays \cite{Pinning}.    
Vortex plastic crystals occur when the temperature is high enough to destroy the
orientational order of the vortex $n$-mers.
Experiments show that as the temperature increases,
certain commensuration peaks disappear,  
which is consistent with the formation of plastic vortex crystals 
\cite{Wu,Pinning}. 

The vortex $n$-merization in honeycomb pinning arrays produces
a novel dynamical symmetry breaking effect 
\cite{Transverse,Transverse2}.  
At $B/B_{\phi} = 2,$ the vortices form dimers with a broken ground state 
symmetry. 
When a longitudinal drive is applied, there is
a positive or negative transverse velocity response
from the moving dimers depending on the direction of the original symmetry
breaking.
At incommensurate fields 
near $B/B_\phi=2$ where dimers are present, 
the ground state is not
globally symmetry broken;   
however, under a longitudinal drive the moving dimers can 
dynamically organize into one of two possible flow states
with a transverse response.    
When the longitudinal drive is held fixed at a constant value, 
the direction of the transverse flow can be switched with 
an externally applied dc transverse drive. 
Both the transverse and longitudinal velocities change abruptly when the
flow direction switches.
If the amplitude of the transverse drive is too weak, 
the system remains locked in the original transverse flow direction. 
When an ac transverse drive is applied in addition to a
fixed dc longitudinal drive, the transverse portion of the vortex flow 
oscillates, resulting in an amplification of 
the transverse response in certain regimes.   
In these regimes, an increase of the transverse ac force 
can cause the longitudinal vortex velocity to decrease. 
The pronounced switching and amplification behaviors of the 
vortex dimers
may be useful for producing transistorlike devices \cite{Toner,Dohmen}.

In this work, we study the vortex dimer response in a sample with a
honeycomb pinning lattice subjected to ac or combined ac and dc drives.
We find switching and jamming phenomena and a rich variety of dynamical
behaviors.
Many of our results should be relevant for systems other than vortices
on periodic pinning arrays.
Simulations, theory, and experiments have shown that
colloidal particles interacting with a periodic substrate 
can exhibit an $n$-merization effect at fillings where there is more than 
one colloid per substrate minimum \cite{Colloid,Brunner,Trizac,Frey}. These states have been termed colloidal molecular crystals and 
have many of the same properties as the 
vortex molecular crystals \cite{Pinning,Transverse,Transverse2}.      
Our results may also apply to other systems such as molecular dimers 
driven over periodic substrates, friction systems \cite{Ying}, or charged
balls on periodic substrates \cite{Guthermann}.    
The symmetry locking effect predicted for vortices moving over periodic arrays
\cite{Nori} has been observed in experiments with colloids moving over 
periodic arrays \cite{Grier}.
We thus expect that the results
described in this work should be generic for colloids on honeycomb arrays.   

\section{Simulation}

We simulate a two-dimensional system of size $L \times L$
with periodic boundary conditions in the $x$ and $y$ directions
containing $N_v$ vortices at a density $n_v=N_v/L^2$. 
The vortex density $n_v$ is proportional to the magnetic field
${\bf B}=B{\hat {\bf z}}$.
Vortices can move under the influence of the Lorentz force from an
applied current ${\bf J}$
or due to interactions with the other vortices. 
In a type-II superconductor the vortex core is not superconducting,
so the vortex motion is overdamped due to dissipation of energy by the core.
The motion of vortex $i$ at position ${\bf R}_i$ 
is governed by the following equation of  
motion:

\begin{equation}
\eta\frac{ d{\bf R}_{i}}{dt} = {\bf F}_{i}^{vv}  + {\bf F}^{p}_{i} + 
{\bf F}^{dc} + {\bf F}^{ac} + {\bf F}^{T}_{i} .
\end{equation}

Here 
the damping constant $\eta=\phi_0^2d/2\pi\xi^2\rho_N$, where 
$d$ is the sample thickness, 
$\phi_0=h/2e$ is the flux quantum, $\xi$ is the superconducting
coherence length, and $\rho_N$ is the normal state resistivity of the
material. 
The repulsive vortex-vortex interaction force is  
\begin{equation}
{\bf F}_{i}^{vv} = \sum^{N_v}_{j\ne i}f_{0}
K_{1}\left(\frac{R_{ij}}{\lambda}\right){\bf {\hat R}}_{ij} , 
\end{equation}
where $K_{1}$ is the modified Bessel function, 
$\lambda$ is the London penetration depth, 
$f_{0}=\phi_0^2/(2\pi\mu_0\lambda^3)$,
the distance between vortices $i$ and $j$ is
$R_{ij} = |{\bf R}_{i} - {\bf R}_{j}|$, and  
${\hat {\bf R}}_{ij} = ({\bf R}_{i} - {\bf R}_{j})/R_{ij}$. 
We measure lengths in units of $\lambda$ and take $L=24\lambda$ so that
the system size is $24\lambda\times 24\lambda$.
Since the vortex-vortex interaction force falls off rapidly with 
$R_{ij}$, we impose a long range cutoff at $7\lambda$.
For very short $R_{ij}$  a cutoff 
is placed at $0.1\lambda$ 
to avoid a divergence in the force.  

The force from the vortex-pinning site interactions is ${\bf F}^{p}_{i}$.
We place $N_{p}$ pinning sites, with a pinning density of $n_{p} = N_{p}/L^2$,
in a honeycomb
array formed by removing every third pinning site from a triangular
array.
The pinning sites have a maximum force of $F_p=0.85$ and a
fixed radius of $r_{p}=0.35\lambda$, which is small enough that only one
vortex is trapped per pinning site. 
Each pinning potential is modeled as an attractive parabolic well, 
giving a force 
\begin{equation}
{\bf F}^{p}_{i} =  \sum_{k=1}^{N_p}\frac{ F_{p}R_{ik}^{(p)}f_0}{r_{p}}
\Theta\left(\frac{r_{p} - R_{ik}^{(p)}}{\lambda}\right) {\bf {\hat R}}^{(p)}_{ik}.
\end{equation}
Here 
$\Theta$ is the Heaviside step function,
${\bf R}_{k}^{(p)}$ is the location of pinning site $k$, 
$R^{(p)}_{ik} = |{\bf R}_{i} -{\bf R}^{(p)}_{k}|$, 
and ${\hat {\bf R}}^{(p)}_{ik} = ({\bf R}_{i}- {\bf R}^{(p)}_{k})/R^{(p)}_{ik}$.  
The matching field $B_{\phi}$ is defined as the field with $N_v=N_p$ 
where there is one vortex per pinning site. 
We fix $B_\phi=0.3125\phi_0/\lambda^2$.
In Fig.~1 we illustrate
the pinning site positions and vortex locations 
for a honeycomb pinning array at 
$B/B_{\phi} = 2.0$.  

\begin{figure}
\includegraphics[width=3.5in]{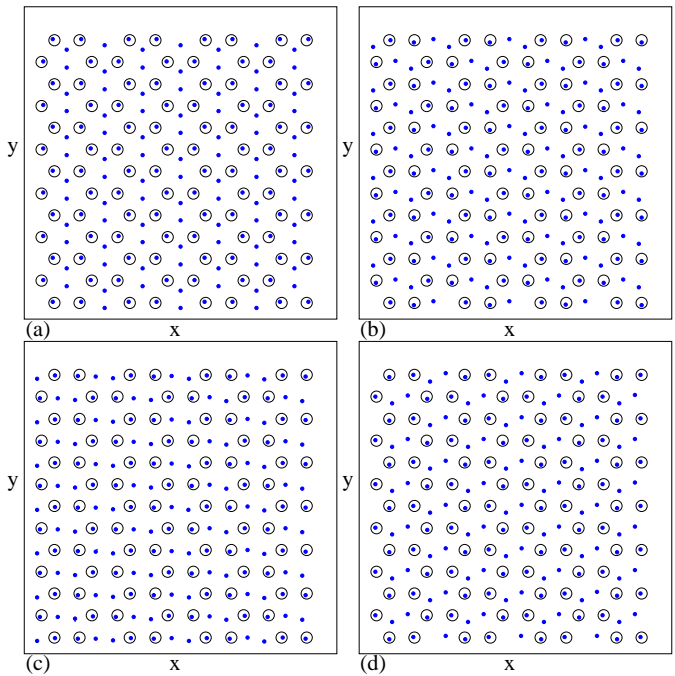}
\caption{
Vortex positions (filled dots) and pinning site locations (open circles)
for a honeycomb pinning array with 
$B/B_{\phi} = 2.0$.
Here the interstitial vortices form 
a dimer state. A circular ac drive is applied with 
$A = 0.075$ and 
$T=3.14\times 10^6$ simulation time steps.
(a) In the initial state at $t=0$ the dimers are aligned 
along the $y$-direction
at $\theta=\pi/2$. 
(b) At
$t/T=0.25$, 
the dimers are tilted 
to $\theta=5\pi/6$.
(c) At $t/T=0.54$, the dimers
are in the process of switching to a new symmetry direction.
(d) At $t/T=0.6$, the dimers have 
switched to 
the orientation 
$\theta=7\pi/6$.
\label{imagefig}
}
\end{figure}

The thermal force ${\bf F}^{T}_{i}$ is modeled 
as random  
Langevin kicks with the correlations 
$\langle{\bf F}^{T}_{i}\rangle = 0$ and
$\langle{\bf F}_{i}^{T}(t){\bf F}_{j}^{T}(t^{\prime})\rangle 
= 2\eta k_{B}T \delta(t-t^{\prime})\delta_{ij}$. 
We obtain the initial vortex configurations by simulated annealing.
In this study we anneal by starting at $f_T=3.0$ and 
spend $5000$ simulation time steps at each
temperature before reducing the temperature in 
increments of $\delta f_T=0.002$. 
Slower annealing times do not change the final $T = 0$ vortex configurations. 

Once the vortex configuration is initialized, 
we 
apply
a uniform dc drive ${\bf F}^{dc}=F^{dc}{\hat {\bf y}}$ 
and ac drive ${\bf F}^{ac}$. 
Unless otherwise noted, the ac force has the form
\begin{equation} 
{\bf F}^{ac} = Af_0\left[\sin(2\pi t/T){\hat {\bf x}} + 
\cos(2\pi t/T){\hat {\bf y}}\right] ,
\end{equation}
with amplitude $A$ and period $T$.
We measure the average vortex velocity response
$V_x=N_v^{-1}\langle\sum_{i=1}^{N_v}{\bf v}_{i}\cdot {\bf {\hat x}}\rangle$
and 
$V_y=N_v^{-1}\langle\sum_{i=1}^{N_v}{\bf v}_{i}\cdot {\bf {\hat y}}\rangle$,
where ${\bf v}_i$ is the velocity of vortex $i$. 
We note that the ability to apply currents in both $x$ and $y$ directions
and measure the voltage response has been demonstrated 
experimentally \cite{Velez1,Sil,Velez2}.    

\section{External Control of the Vortex Dimer Orientation}

We first consider a honeycomb pinning lattice
at $B/B_{\phi} = 2.0$ 
with a circular ac drive and no dc drive. 
At this matching field, the interstitial vortices 
form dimer states in the large interstitial regions of the honeycomb pinning
lattice, as shown in Fig.~1(a) and studied previously in Ref.~\cite{Pinning}. 
The dimers have an effective ferromagnetic coupling
and are all aligned in the same direction. 
In previous work we showed that in the absence of a drive,
the ground state configuration of the dimers can be 
aligned in one of three directions \cite{Pinning}. 
Here we show that an  
external ac drive can couple to the 
dimers and change their orientation
provided that the ac amplitude is small enough that the interstitial
vortices do not depin. 

In Fig.~1
we illustrate the time evolution of a system under an ac drive
with $A=0.075$ and 
$T=3.14\times 10^6$ simulation time steps.
The dimers are initially aligned in the $y$-direction as shown in Fig.~1(a).
We denote the direction of the dimer orientation as $\theta$, measured
on the unit circle. 
Here, the initial orientation is $\theta=\pi/2$.
The ac drives induces two types of dimer motion. 
The first is a rotation of the dimers in phase with the ac drive and
the second is an abrupt {\it switching} where 
the dimers suddenly change their orientation to a new symmetry direction.
We characterize these different types of motion by 
examining the vortex positions and trajectories as well as 
the time traces of $V_x$ and $V_y$.
Fig.~1(b) shows a snapshot of the dimer positions at $t/T=0.25$ when
the dimers have tilted to align at 
$\theta=5\pi/6$.
At $t/T=0.54$ the dimers are in the process of switching between 
$\theta=5\pi/6$ and $\theta=7\pi/6$,
and align as shown in Fig.~1(c) to form an overall 
stripelike pattern that incorporates the pinned
vortices.
In Fig.~1(d) at $t/T=0.6$,
the dimers have 
switched to an alignment 
of $\theta=7\pi/6$.
The time sequence shows that the dimers rotate in phase
with the circular
ac drive.

In Fig.~2 we plot the vortex trajectories 
corresponding to the system shown in Fig.~1. 
The first two switches in the cycle (during a time interval of $0.6T$) 
appear in Fig.~2(a), while
Fig.~2(b) indicates that the dimers move in a closed triangular
orbit during a complete cycle.
A combination of smooth and switching dimer motion appears in $V_x$ and
$V_y$ as a function of time in Fig.~3.
The sinusoidal envelopes of the velocity traces are interlaced with 
spike features formed when the dimers abruptly switch to a new
orientation.
The first jump at $t/T=0.2$ occurs at the transition from the
$\theta=\pi/2$ orientation shown in Fig.~1(a) to 
the $\theta=5\pi/6$ orientation shown in
Fig.~1(b).
The lower vortex in each dimer shifts
in the positive $x$ direction with a smaller displacement 
in the negative $y$ direction, while the other vortex in the dimer 
shifts in the negative $y$ direction with a smaller displacement in the
negative $x$ direction. 
The sum of these motions gives a positive $V_x$ pulse and a simultaneous
negative $V_y$ pulse in Fig.~3.
In the next switching event at $t/T=0.54$,
the upper vortex in each dimer shifts
in the negative $y$ direction while the other dimer 
vortex shifts in the positive
$y$ direction, reducing the net value of $V_y$.
At the same time, both vortices in the dimer move in the negative
$x$ direction, enhancing the magnitude of $V_x$ as shown in Fig.~3.
This process repeats each period.
The switching behavior occurs since the vortex dimers sit
in a caging potential with a sixfold modulation created by 
the repulsive interactions with the six vortices located 
at the neighboring pinning sites. 
This modulation provides an energy barrier for the switching of the dimer from
one orientation to another.
The field-induced switching  effect we observe
is similar to liquid crystal behavior 
where the orientation of rod shaped molecules 
can be altered with external fields. 
Thus, we term the dimer switching a polarization effect, and note that 
it could be used to create numerous novel vortex devices.
For example, 
the orientation of the dimers can be rotated in 
such way as to maximize the critical current
in certain directions. 
Additionally, logic states could be assigned with 
different logic values corresponding to different dimer
orientations.   

\begin{figure}
\includegraphics[width=3.5in]{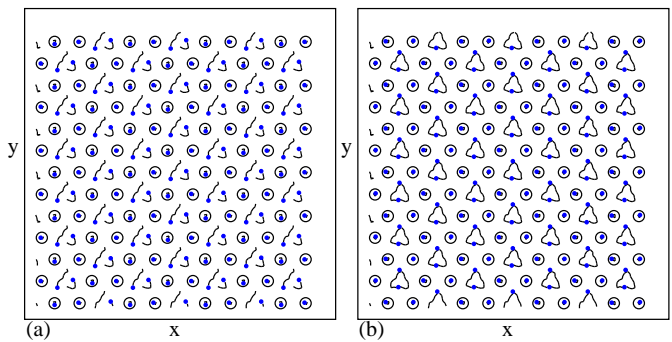}
\caption{
Vortex positions (filled dots), pinning site locations (open circles), 
and vortex trajectories (lines)
for the same system in Fig.~1. 
(a) The trajectories during a time interval of 
$0.6T$ when two switching events occur.
(b) The  trajectories during a time interval of 
$1.1T$ showing the complete vortex
orbits. The interstitial vortices follow the same orbits 
for all later times.
}
\end{figure}

\begin{figure}
\includegraphics[width=3.5in]{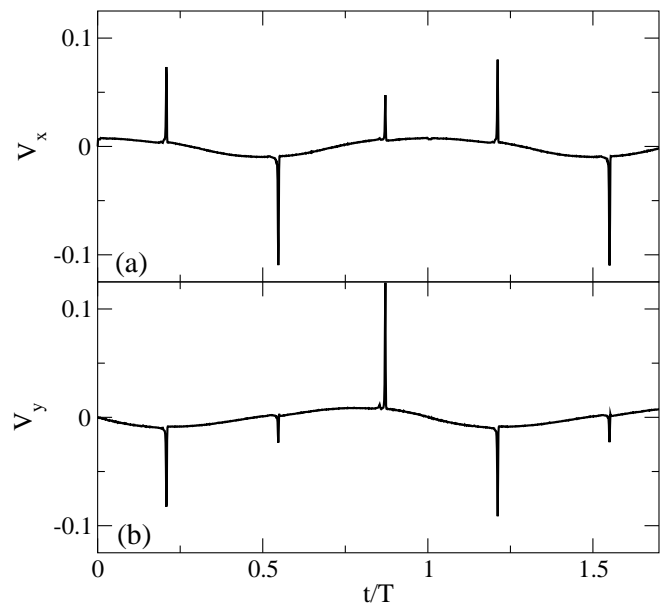}
\caption{
The instantaneous vortex velocity vs time for the system in Fig.~1. 
(a) $V_x$. (b) $V_y$. The spikes correspond to the abrupt switching events 
where the dimers change orientation.   
}
\end{figure}

The pinned switching behavior shown in Fig.~3 only 
occur under certain conditions. If the ac amplitude $A$ is small, the
dimers are unable to switch and remain locked in a single orientation.
Conversely, if the ac amplitude is too large, the dimers 
escape from the interstitial caging potential and channel along 
one of six angles: $\theta=\pi/6$, $\pi/2$, $5\pi/6$, $7\pi/6$,
$3\pi/2$, or $11\pi/6$.
In Fig.~4(a,b) we plot $V_x$ and $V_y$ versus time
for the same system in Fig.~3 
but with a smaller ac amplitude of $A=0.03$ which 
is too weak to induce switching. The velocity 
curves are smooth and merely follow the ac input drive as the vortices
shift slightly in the interstitial sites. 
When the external drive is large enough, the dimers depin from the
large interstitial
regions and channel along 
one of the six easy-flow directions of the lattice, giving
the velocity response shown in Fig.~4(c,d) for $A=0.125$. 
The channeling effect is very pronounced 
due to the
significant barriers that the vortices must overcome in order
to lock their motion into a new easy-flow direction.  In fact, the 
locking is so strong that when the vortices switch 
to a different flow direction  they
skip over every other easy-flow direction, 
so that only three of the possible
six symmetry flow directions are followed in a given realization.
Which of the two sets of three flow 
directions that the system chooses depends on the
initial conditions. 
In Fig.~4(c,d) at $A=0.125$, the dimers initially 
flow along $\theta=\pi/2$,
similar to the 
motion shown in Fig.~5(a). 
Figure~4(c,d) indicates that during the first portion of the drive cycle,
$-0.15\leq t/T<0.15$, 
$V_y$ is finite and $V_x$ is zero, as expected for motion that is locked in
the $y$ direction.
For $0.15 \leq t/T < 0.5$, 
the $x$ component of the ac drive becomes
strong enough to reorient the dimers so that they 
channel along a different symmetry direction.
The vortices lock to 
$\theta=11\pi/6$ and have the same type of motion illustrated in Fig.~5(e).
During $0.5 \leq t/T < 0.85$, the vortex motion switches and
flows along $\theta=7\pi/6$.
Finally the vortex motion returns to $\theta=\pi/2$.  In this cycle, the
vortices never flow along $\theta=\pi/6$, $3\pi/2$, or $5\pi/6$. 
By changing the initial conditions slightly, we can obtain
vortex motion along those three angles only (and not along $\pi/2$, $11\pi/6$,
or $7\pi/6$) with a fifty percent probability.
For $A=0.125$, there are small windows of pinned phases 
between the sliding locked phases.  
In Fig.~6 we plot
the vortex trajectories during an entire drive period 
showing the triangular pattern formed as the vortices flow along three
different directions.

\begin{figure}
\includegraphics[width=3.5in]{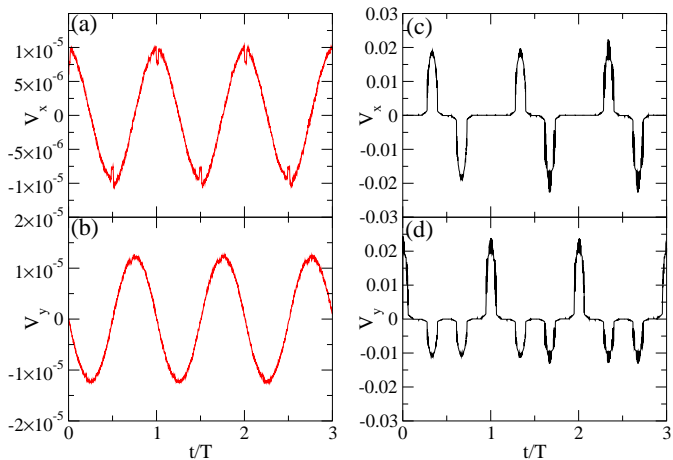}
\caption{(a) $V_x$ and (b) $V_y$ versus time
for the same system in Fig.~1 with $A = 0.03$ 
where the dimers do not switch. 
(c) $V_x$ and (d) $V_y$ for the same system with 
$A = 0.125$ where the ac amplitude is high enough to induce a 
sliding of the dimers. In each driving period, 
the dimers initially slide 
along $\theta=\pi/2$ in the positive $y$-direction, become pinned,  
and then slide along 
$\theta=11\pi/6$ and $\theta=7\pi/6$ in turn.
}
\end{figure}
\begin{figure}
\includegraphics[width=3.3in]{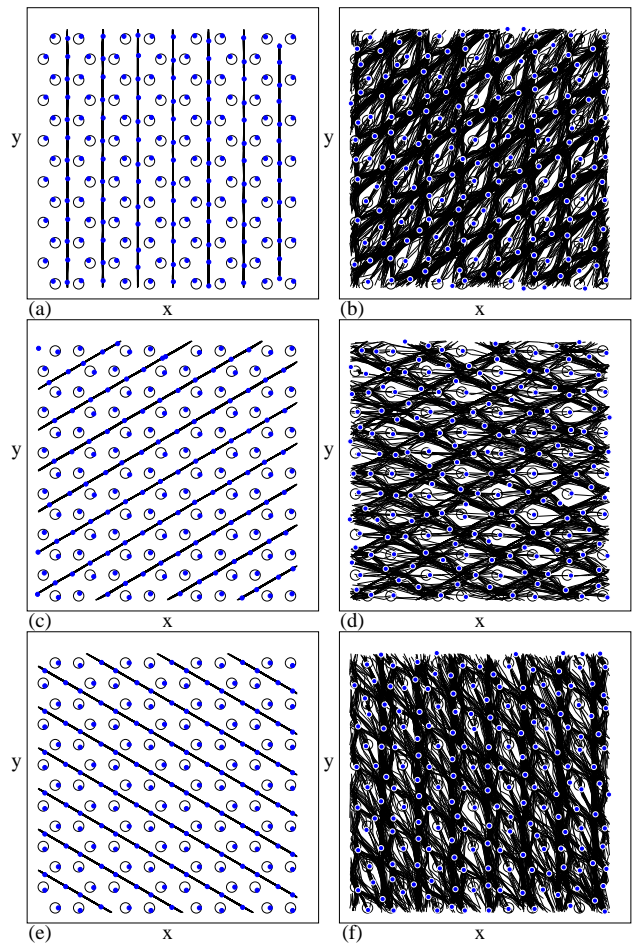}
\caption{
Vortex positions (filled dots), pinning site locations (open circles), 
and vortex trajectories (lines) for the system in Fig.~1 at $A=0.35$ during
half of a drive cycle.
The points from which the images are taken are marked with bold letters in
Fig.~8(a).
(a) The dimers initially move in the $y$-direction, $\theta=\pi/2$. 
(b) Transition to the disordered flow state in which a portion of the pinned
vortices depin.
(c) The dimers lock to $\theta=\pi/6$.
(d) Transition to another disordered flow state where a portion of the pinned
vortices depin; the average velocity is in the $x$ direction.
(e) The dimers lock to 
$\theta=11\pi/6$.
(f) Transition to a disordered flow state in which a portion of the pinned
vortices depin.
The remainder of the drive cycle has the same appearance as (a-f) but with the
motion of the vortices reversed to follow $\theta=3\pi/2$, 
$7\pi/6$, and $5\pi/6$.
}
\end{figure}

\begin{figure}
\includegraphics[width=3.5in]{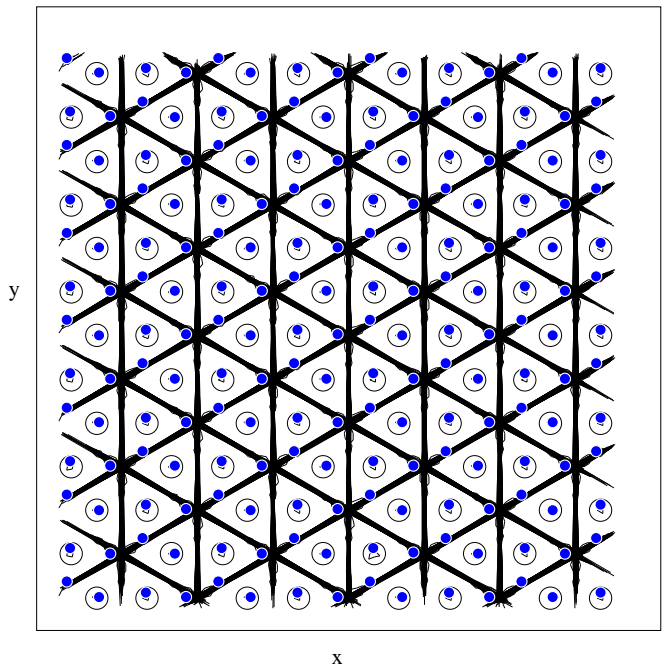}
\caption{ Vortex positions (filled dots), pinning site locations (open circles),
and vortex trajectories (lines) for the system in Fig.~4(c,d) at $A = 0.125$
during a complete drive cycle. The dimers slide along three different angles 
so the trajectories form a triangular pattern.   
}
\end{figure}

For ac amplitudes 
$0.11 < A \leq 0.2$, Fig.~7(a) shows that 
only three of the six possible flow directions appear and that 
$V_x$ and $V_y$ simultaneously
drop to zero between the sliding regimes in each period, indicating the
existence of a temporary pinned phase.
For  $A > 0.2$    
the system passes directly from 
one siding state to another without pinning, as indicated in Fig.~7(b), and
flow occurs in all six possible directions.
Figure~7(b) shows that at $A=0.3$, $V_x$ and $V_y$ have discontinuities when the
system switches from one flow direction to another, but that
$V_x$ and $V_y$ are never simultaneously zero.
Within a given flowing locked phase, $V_x$ and $V_y$ are not fixed but can show
considerable variations.

\begin{figure}
\includegraphics[width=3.5in]{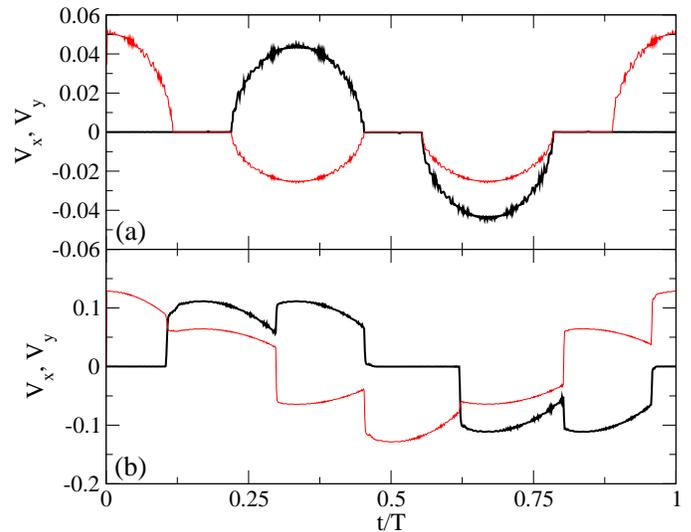}
\caption{
$V_x$ (heavy line) and $V_y$ (light line) versus time for the system in
Fig.~1. (a) At $A = 0.16$ there are three pinned phases 
between the three sliding phases in each
period.   
(b) At $A = 0.3$, the dimers are always sliding along the 
different symmetry directions
and there are no pinned phases.
}
\end{figure}

For $0.2 < A < 0.335$, only the interstitial vortices are moving, while
for
$A \geq 0.335$ a portion of the vortices located at  
the pinning sites become mobile during
certain parts of the ac drive cycle. 
This produces
additional disordered flow phases between the flowing locked phases.
We plot $V_x$ and $V_y$ during one drive cycle at $A=0.35$ in Fig.~8(a) and
show the corresponding vortex trajectories in Fig.~5.
At the beginning of the drive cycle, the  
dimers are locked to flow along $\theta=\pi/2$, as shown
in Fig.~5(a). 
The system does not 
jump directly into an ordered channel flow of dimers along $\theta=\pi/6$,
but instead first passes through an 
intermediate flow state near point {\bf b} in Fig.~8(a). 
The vortex trajectories are disordered, 
as shown in Fig.~5(b). 
Within this regime a portion of the vortices 
at the pinning sites depin due to the
combination of the vortex-vortex interaction 
forces and the applied drive. 
The result is a random flow 
which appears as enhanced fluctuations in $V_x$ and $V_y$.
From the random phase, the system reorders into 
a locked flowing state along $\theta=\pi/6$ illustrated in Fig.~5(c),
where all the pinning sites are again occupied and 
the dimers move in an ordered fashion.  
The system then enters another disordered flow regime shown in 
Fig.~5(d). 
The transition into this disordered phase produces a jump in 
$V_x$.
As the ac drive period progresses, the vortices lock into another flowing
state along $\theta=11\pi/6$ illustrated in Fig.~5(e), 
followed by the disordered flow phase shown in Fig.~5(f). 
During the second half of the drive cycle, a similar series of ordered and
disordered phases occur as the vortices successively lock into $\theta=3\pi/2$,
$\theta=7\pi/6$, and $\theta=5\pi/6$, before finally returning back to the
starting point of locked flow along $\theta=\pi/2$.
In all cases, the random flow phases are associated with increased
fluctuations of $V_x$ and $V_y$.
The vortex velocity distributions are broadened in 
the random flow phase, whereas in the ordered flow phases the velocities
are synchronized.        

\begin{figure}
\includegraphics[width=3.5in]{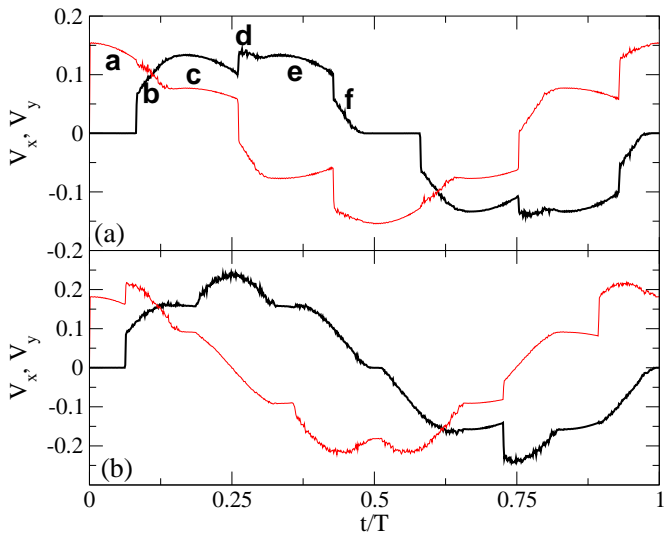}
\caption{
$V_x$ (heavy lines) and $V_y$ (light lines) versus time
for the same system in Fig.~1. 
(a) At $A = 0.35$, a portion of the vortices at the pinning sites 
start to depin and generate random phases between the locked sliding phases. 
The labels {\bf a}, {\bf b}, {\bf c}, {\bf d}, {\bf e}, and {\bf f} 
correspond to the portions of the period illustrated in
Fig.~5.
The disordered flow phases appear at {\bf b}, {\bf d}, and {\bf f} 
during the transitions
between different locked sliding states. 
(b) At $A = 0.405$, the random flow phases are broader.
}
\end{figure}

In Fig.~8(b) we plot $V_x$ and $V_y$ versus time for $A = 0.405$ where 
only a small window of each locked sliding phase 
occurs between the random flow regimes.
For $A>0.42$, the locked sliding phases 
vanish and the flow is disordered throughout the ac drive cycle, 
although some slight channeling persists within the random flow phase.      

We summarize
the results for $0 < A < 0.45$ in a dynamic phase diagram in Fig.~9,
where we plot the phases for $A$ versus time during half an ac drive period. 
For $A < 0.05$ we find a pinned locked (PL) phase in which 
all the vortices remain pinned and the dimers
do not show any switching behavior. 
The dashed line at $A=0.05$ separates the PL  from the pinned switching (PS)
regime where the dimers remain pinned but change orientation.
For  $0.11 < A < 0.16$ the PS phase is lost over much of the drive cycle and is 
replaced by a series of sliding states along every other symmetry angle.
Here, the vortices follow $\theta=\pi/2$ and $\theta=11\pi/6$ 
in this half of the
drive cycle, separated by the remnants of the PS phase.
The PS regime disappears completely for $A\geq 0.16$ and 
for $0.16 \leq A < 0.3$, the vortices lock into each
symmetry angle in turn, passing from $\theta=\pi/2$ to $\theta=\pi/6$ 
to $\theta=11\pi/6$, and so forth. 
For $0.325 \leq A < 0.435$, the sliding phases are separated 
by intrusions of the random flow (R) phase
which increase in extent until  
the sliding states are completely lost for $ A \geq 0.435$.          

\begin{figure}
\includegraphics[width=3.5in]{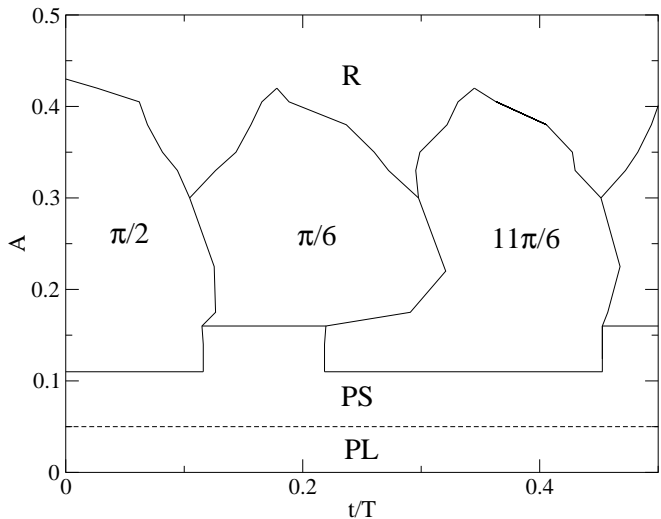}
\caption{
The dynamic phase diagram for the system in Fig.~1 
showing ac amplitude $A$ vs time.
For low $A$ the dimers are in the pinned locked (PL) state.
For $0.05 < A \leq 0.11$ the dimers are in the pinned switching (PS)
state where they remain pinned but switch between
different orientations as shown in Fig.~2 and Fig.~3. 
The dashed line separates the 
PL and PS phases.
For $0.11 < A < 0.16$, the vortices pass through a series of 
sliding phases along $\theta=\pi/2$, $\theta=\pi/6$, and $\theta=11\pi/6$.
The sliding phases are separated by pinned phases 
for $0.11 < A < 0.16$ and by random flow (R) phases for $0.325 < A < 0.435$.
}
\end{figure}

\subsection{High ac Amplitude and Order-Disorder Transitions}

The vortices depin and flow in a random phase when the ac amplitude is large
enough to overcome the pinning force.
For even higher ac drive amplitudes, we find that
the system enters a new dynamical regime.  The vortex flow changes from  
the channeling of interstitial vortices between occupied  
pinning sites to a regime where all the vortices take part in the motion and 
a portion of the vortices move in locked channels that pass
{\it through} the pinning sites. 
For ac amplitudes just above $A=0.435$ where the dimer channeling is lost, 
the system is disordered and the vortices undergo plastic flow in which
some vortices are temporarily trapped at the pinning sites.  
The features in the velocity versus time curves are washed out and 
we find only a weak channeling of the vortex motion. 
In this regime, the vortex lattice is heavily defected. 
As the ac amplitude 
is further increased, the effectiveness of the pinning
diminishes and the vortices begin to partially reorder along  
certain symmetry directions of the pinning lattice. 

To quantify the reordering we use a
Voronoi construction to measure the fraction of six-fold 
coordinated vortices $P_{6}=N_v^{-1}\sum_{i=1}^{N_v}\delta(z_i-6)$, 
where $z_i$ is the coordination number of vortex $i$. 
For a perfectly triangular vortex lattice, $P_6=1$.
In Fig.~10(a) we plot  $V_{x}$ and $V_{y}$ vs time for 
$A = 0.43$, where the vortices are in the random flow (R) regime.
Small features in $V_x$ and $V_y$ indicate the persistence of
partial channeling even in the random flow phase.
Figure 10(b) shows the corresponding measurement 
of $P_{6}$, 
which has an average value of $P_6 \approx 0.5$ due to the
highly defective nature of the vortex lattice in the random flow phase. 
At $A=0.5$, Fig.~10(c) indicates that the channeling features 
disappear and are replaced by
new steplike features in
$V_x$ and $V_y$. 
Figure 10(d) shows that the steps are
correlated with six maxima in 
$P_{6}$, 
indicating that the vortices are partially ordered along the 
steps and that a different type of
channeling effect is occurring. 
The overall vortex lattice remains disordered throughout the 
drive cycle; however, when the drive falls on a step,
$P_{6} \approx 0.68$, while between the steps, $P_{6} \approx 0.5$. 

\begin{figure}
\includegraphics[width=3.3in]{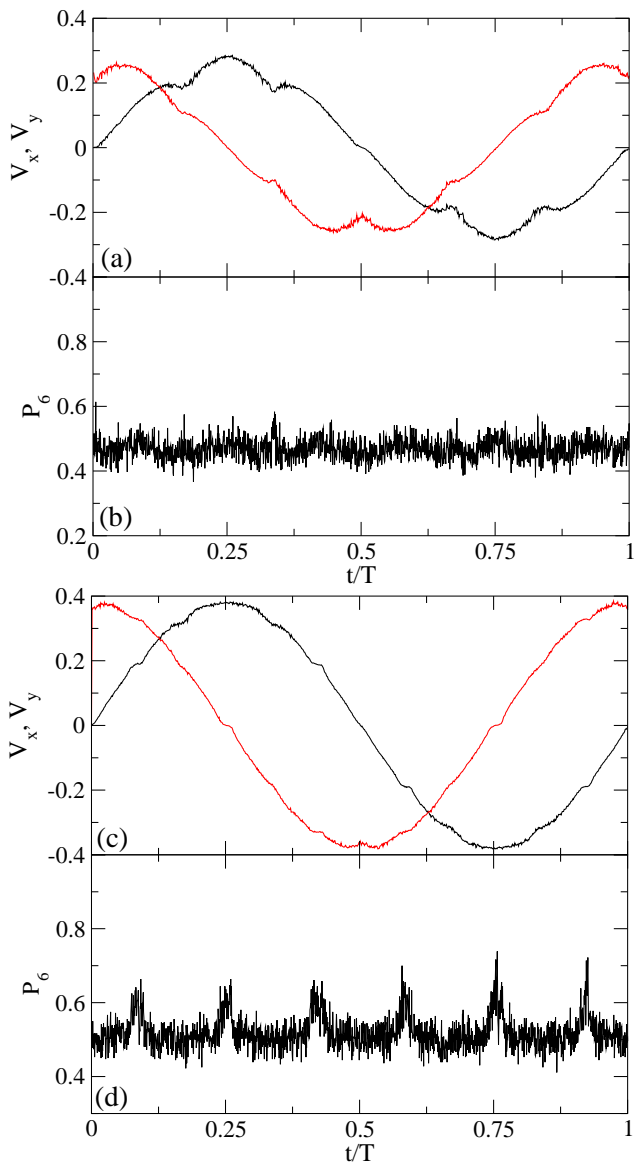}
\caption{
(a) $V_x$ (dark line) and $V_y$ (light line) versus time for the system in
Fig.~9 at $A=0.43$, where the flow is disordered.
Although the system is in the random phase, there is still some 
mild channeling of the vortices which appears as 
modulations of 
$V_x$ and $V_y$.
(b) The corresponding fraction of six-fold coordinated vortices $P_{6}$
vs time. For a completely triangular
lattice $P_{6} = 1.0$. 
Here $P_{6} \approx 0.5$, indicating that the vortex lattice is
disordered throughout the whole driving cycle. 
(c) $V_x$ (dark line) and $V_y$ (light line) versus time for the system
in Fig.~9 at $A = 0.5$.  The channeling effects are completely destroyed;
however, new steplike features are beginning to appear.
(d) The corresponding $P_{6}$ vs time
shows enhancements that are correlated with the
steplike features in $V_x$ and $V_y$, 
indicating that the vortices are gaining some partial ordering on the steps.
}
\end{figure}

At $A=0.75$, shown in Fig.~11(a), the steps in $V_x$ and $V_y$ are much denser
and are associated with a series of peaks and dips in $P_6$, shown in 
Fig.~11(b).
Between the steps, $P_6\approx 1$, indicating a triangular vortex lattice,
while on the steps, the vortices are less ordered and $P_6\approx 0.68$.
The changes in $P_6$ 
indicate that the vortices undergo a series of 
order-disorder transitions during the ac drive cycle.
The average minimum amount of vortex order present at $A = 0.75$
is larger than average maximum vortex order at $A=0.5$, as indicated by 
comparing $P_6$ in Fig.~11(b) and Fig.~10(d).
In Fig.~12(a) we show a blowup of $V_x$ and $V_y$ for
$0.05 < t/T < 0.2$ from Fig.~11(a) to highlight the order-disorder 
transition that occurs when the vortices 
enter and exit a single velocity step. 
We plot the corresponding $P_6$ values in Fig.~12(b). 
Within the nonstep intervals, where $V_x$ and $V_y$
are smoothly increasing or decreasing, $P_{6} \approx 1$, 
indicating that the vortices are mostly sixfold coordinated. 
The dashed lines in Fig.~12 outline the step located at
$0.072 < t/T < 0.102$. 
A dip in $P_6$ occurs as the system both enters and exits this step,
indicating enhanced disorder in the vortex lattice.
On the velocity step, $P_{6}$ increases to $P_6 \approx 0.85$ but
the vortex lattice structure is not completely ordered. 
Outside of the step, the vortices regain a nearly triangular ordering
until reaching the next velocity step at $t/T \approx 0.15$.
In Fig.~13 we plot the vortex trajectories for the system in 
Fig.~12 before, during, and after the step, where the step onset 
$t_{step}/T=0.072.$
Below the step, Fig.~13(a) shows that the vortex trajectories are disordered, 
while on the step Fig.~13(b) shows more ordered vortex motion 
with a partial locking to $\theta=\pi/3$.  The vortices channel along the 
pinning sites rather than between them.
Above the step, the vortex motion is again disordered, as illustrated in 
Fig.~13(c). 

\begin{figure}
\includegraphics[width=3.5in]{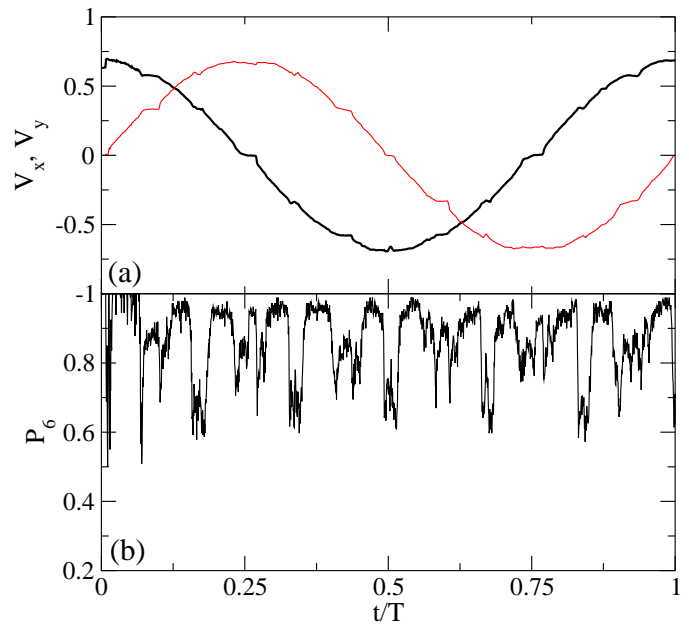}
\caption{(a) $V_x$ (dark line) and   
$V_y$ (light line) vs time for the same system in Fig.~10 
with a higher ac amplitude of $A = 0.75$. (b) The corresponding
$P_6$ vs time shows a series of order to disorder transitions.  
}
\end{figure}

\begin{figure}
\includegraphics[width=3.5in]{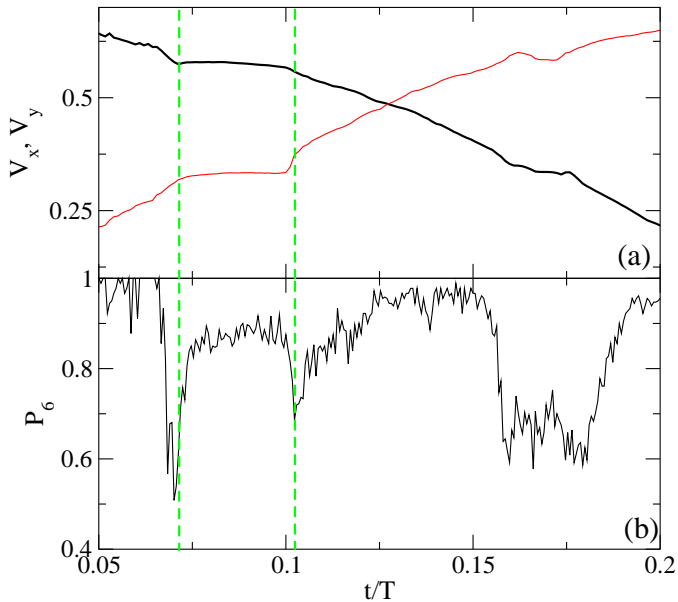}
\caption{A blowup of the region $0.05 < t/T < 0.2$ from Fig.~11 
with $A=0.75$
highlighting the order to disorder transitions across a step.
(a) $V_x$ (dark line) and $V_y$ (light line) versus time.
(b) The corresponding $P_6$ vs time.
The dashed lines highlight the transitions into and out of the step;
the vortex positions are disordered at the transitions and more ordered
on the step.
}
\end{figure}

\begin{figure}
\includegraphics[width=3.5in]{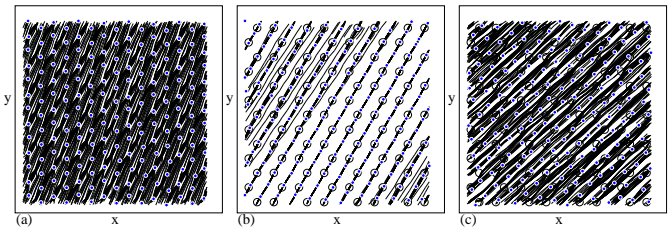}
\caption{
Vortex positions (filled dots), pinning site locations (open circles),
and vortex trajectories (black lines)  for the system in Fig.~12.
(a) $(t_{step}-t)/T=0.06$, below the locking step.  The vortices are
initially not locked to a specific angle. 
(b) In the center of the step region, $(t-t_{step})/T=0.014$, the
vortices channel along $\theta=\pi/3$.
(c) Above the locking step, $(t-t_{step})/T=0.06$,
the vortex trajectories are disordered.
}
\end{figure}

The other velocity steps in Fig.~11(a) also correspond 
to regimes where the vortices 
channel along the pinning sites at different angles. 
This type of channeling motion 
along pinning sites was studied previously for
vortices in square pinning arrays at fields close to the 
matching field $B/B_{\phi} = 1.0$ \cite{Nori}.
In that system, a fixed dc drive $F^x_d$ was applied in the longitudinal
direction and a transverse dc drive $F^y_d$ was gradually increased.
A series of steps in the transverse velocity appeared, forming
a devil's staircase structure. 
The velocity steps form when the vortex motion is
locked to certain angles with respect to the symmetry axis of 
the pinning array.
Specifically, for a square array steps form
at drive ratios of  $F^{y}_{d}/F^{x}_{d} = m/n$, 
where $m$ and $n$ are integers.  
For a triangular pinning array,
velocity steps occur when
$F^{y}_{d}/{F^{x}_{d}} = \sqrt{m}/{2n +1}$. 
For the honeycomb array, we expect a similar set of
velocity steps to occur at the same drive ratio values as in
the triangular array, but unlike the triangular array, the steps should be
particularly pronounced when the drive angle is along $\theta=\pi/3$,
$\theta=\pi/2$, $\theta=2\pi/3$, or the complements of these angles.
In Ref.~\cite{Nori}, 
a series of order-disorder transitions were observed that are similar
to the order-disorder transitions we find in the honeycomb array.

In Fig.~14(a) we plot the Voronoi construction of a snapshot of the vortex
configuration at $(t_{step}-t)/T=0.06$, below the step highlighted in Fig.~12.
In the non-step regions,  
the vortex lattice is almost completely triangularly ordered 
and contains only a small number of $z_i=5$ and $z_i=7$ paired defects. 
In Fig.~14(b) at $t=t_{step}$,
the vortex lattice
becomes strongly disordered and there is 
a proliferation of $z_i=5$ and $z_i=7$ defects.
In the center of the velocity step at $(t-t_{step})/T=0.014$, Fig.~14(c) shows
that the density of defects drops
compared to Fig.~14(b).
The remaining $z_i=5$ and $z_i=7$ defects are aligned in the direction of the
channeling motion, which can be seen from the trajectory plots in Fig.~13(b). 
The defect alignment indicates that on the steps, the vortex lattice has the
characteristics of a moving smectic.
Above the step, as shown in Fig.~14(d) at $(t-t_{step})/T=0.06$, the vortices
reorder into a nearly triangular lattice.
A similar set of vortex lattice structures appear for each step and the
topological defects align in the channeling direction on the steps.
At the transitions on and off of each step, the vortex lattice disorders
when a portion of the vortices move in a channeling direction while another
portion of the vortices move in the driving direction.
This combination of vortex motion disorders the lattice.

We note that in general we do not observe {\it complete} 
locking, and even on a step, not all the vortices align with the drive
as shown in Fig.~13(b). 
Much stronger velocity steps with complete locking appear for 
vortex motion in  square pinning arrays 
at lower vortex densities of $B/B_{\phi} = 1.05$, as shown in Ref.~\cite{Nori}.
For the system in Fig.~13, $B/B_\phi=2$ and the locking effect is reduced
since there are twice as many vortices as pinning sites.
The vortex-vortex interactions tend to cause the 
vortex lattice to be triangular. 
The higher order velocity steps are small.
In some cases, 
a high-order step disappears
when the vortex-vortex interactions dominate to create 
a floating moving solid where the vortices form a triangular lattice        
and no locking effect occurs.
In general we expect the steps to be more pronounced at
lower vortex densities. 
For higher vortex densities, only the velocity steps corresponding to the
dominant symmetry directions of the pinning lattice will be discernible.

\begin{figure}
\includegraphics[width=3.5in]{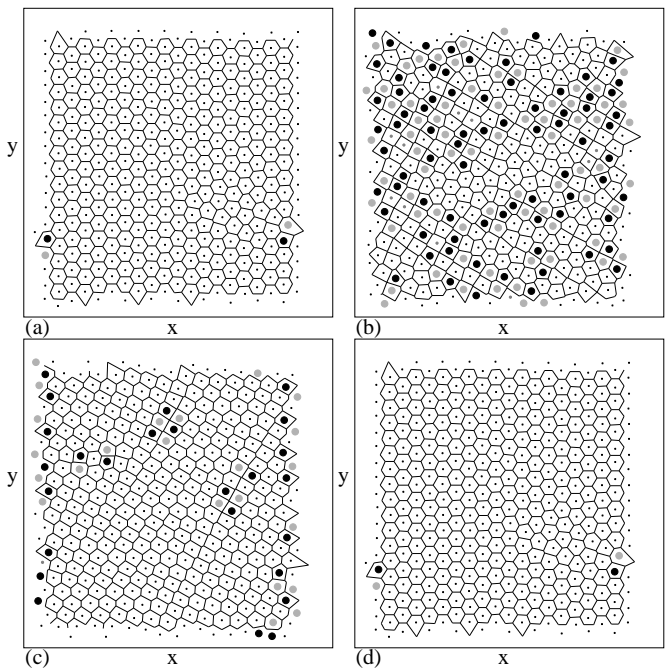}
\caption{
Voronoi constructions of vortex positions
for the system in Fig.~13.  White polygons are sixfold
coordinated, $z_i=6$; large black dots indicate polygons with $z_i=5$
and large gray dots denote polygons with $z_i=7$. 
(a) Vortex configuration below the locking step at 
$(t_{step}-t)/T=0.06$ at the point shown in 
Fig.~13(a).
The vortex lattice is mostly triangular. 
(b) Vortex configuration at $t=t_{step}$, 
the transition into the locking phase. 
There is a proliferation of topological defects. (c) 
Vortex configuration at $(t-t_{step})/T=0.014$ in the locking phase illustrated
in Fig.~13(b).
There is some partial ordering of the lattice, and many of the
$z_i=5$ and $z_i=7$ defect pairs 
are aligned in the direction of the locked motion. 
(d) Vortex configuration at $(t-t_{step})/T=0.06$ above the locking step 
as shown in Fig.~13(c).  The vortex lattice is ordered.
}
\end{figure}

\subsection{Vortex Velocity Evolution with ac Amplitude} 

As the ac amplitude increases, the velocity of the vortices increases
and the locking step intervals diminish in width.
Additionally, the number of steps that generate vortex disorder at their edges
decreases, and for
higher drives the disordering transitions occur 
only along the six prominent locking directions, as 
shown in Fig.~15 for $A = 5.0$. 
In Fig.~16 we compare the vortex lattice ordering at 
$A = 0.5$ and $A = 5.0$ by simultaneously plotting
$P_{6}$ versus time for these two ac drive amplitudes. 
The change in lattice order when the vortices move along a symmetry direction
of the pinning lattice is clear.
At $A = 0.5$, the vortices are {\it more} ordered when moving along a symmetry
direction,
while at higher ac amplitudes this reverses 
and the vortices are {\it less} ordered when 
they move along a pinning symmetry direction. 

\begin{figure}
\includegraphics[width=3.5in]{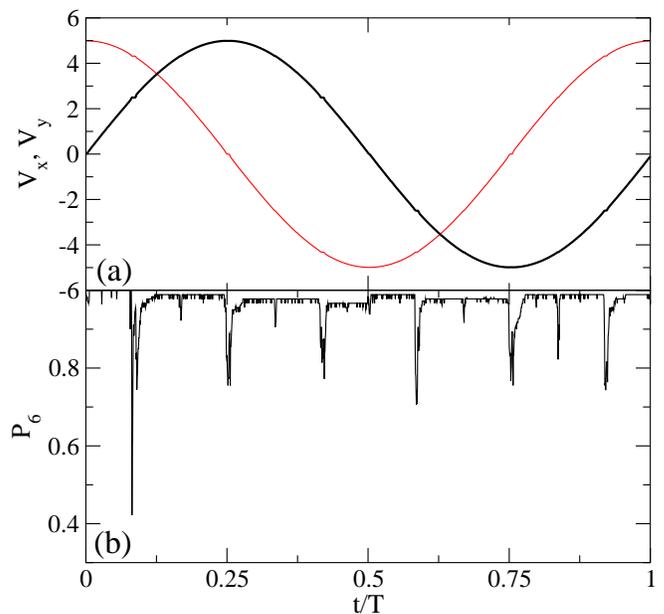}
\caption{
(a) $V_x$ (dark line) and $V_y$ (light line) versus time
for the same system in Fig.~10 with $A = 5.0$. 
Here the step structures are very small. 
(b) The corresponding $P_6$ versus time.
Disordering transitions occur at the steps which are associated with
the six symmetry angles of the pinning lattice, 
where some locking motion occurs.  
}
\end{figure}

\begin{figure}
\includegraphics[width=3.5in]{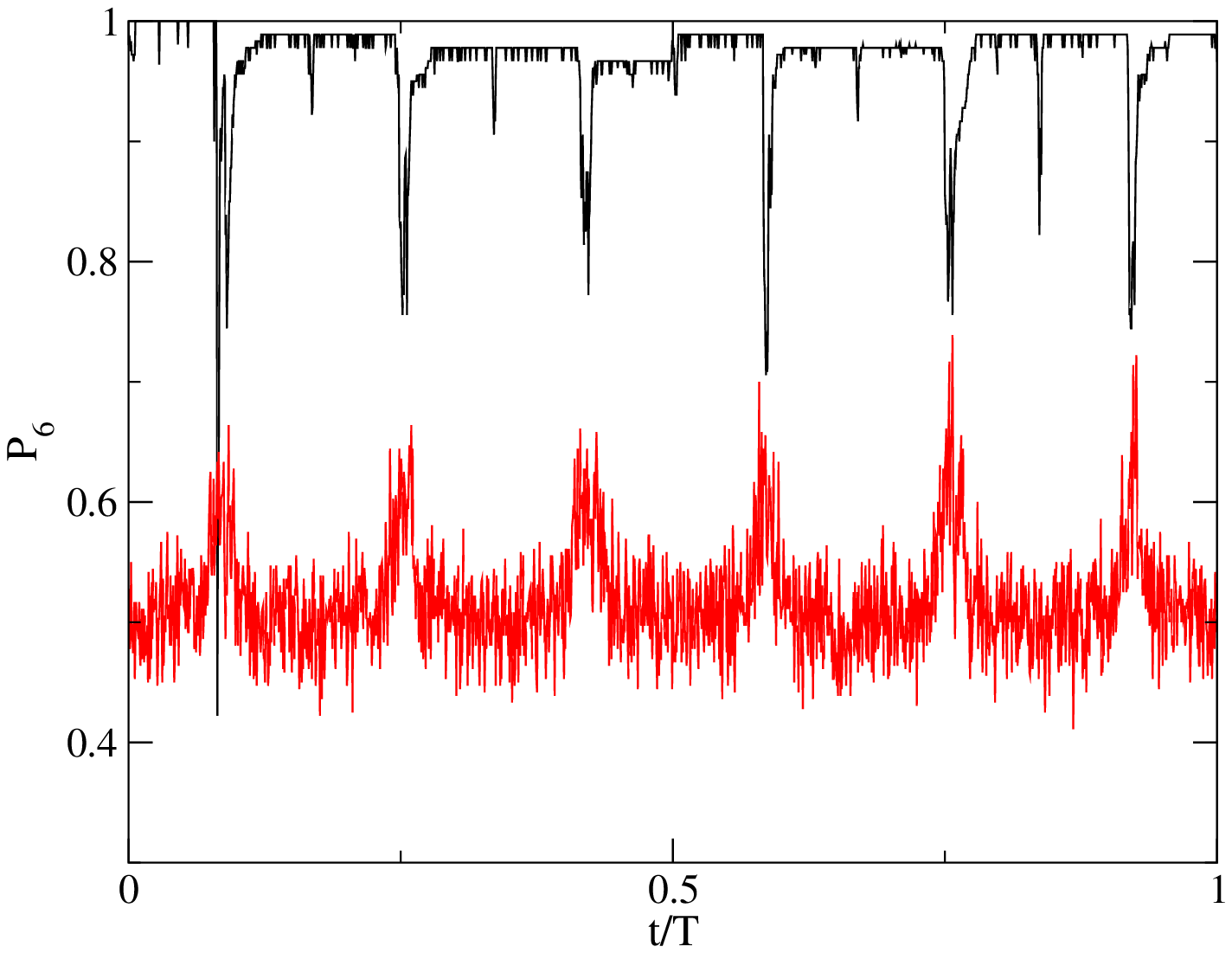}
\caption{
$P_6$ versus time 
for the system in 
Fig.~10 comparing the random phase at $A = 0.5$ (lower curve), where the 
vortices are more ordered along the six symmetry directions, to the
phase at $A = 5.0$ (upper curve), 
where the system is most disordered when sliding along the 
six symmetry directions of the pinning lattice.
}
\end{figure}

The overall change in $V_x$ versus time for varied $A$ is illustrated in
Fig.~17.
In Fig.~17(a) we plot $V_{x}$ during the first half of the period for 
$0.16 \leq A \leq 0.43$. 
One of the most striking features is that near $t/T=0.25$ 
for $ A = 0.35$,
which is highlighted by the darker curve, 
a new maximum in $V_x$ develops that is associated with the random 
flow.  This maximum becomes centered at $t/T=0.25$ 
and increases in amplitude as $A$ increases.
An increasing fraction of the vortices depin during the random flow phase
as $A$ increases, and thus the magnitude of $V_x$ increases.
The onset of the random phase is correlated with 
an increase in the level of fluctuations in $V_x$.
For $A\geq 0.435$, there are no longer any locked sliding regimes;
however, a small amount of locking persists and produces shoulders
in $V_{x}$ near $t/T=0.15$ and $t/T=0.35$ 
that gradually wash out at higher values
of $A$. In Fig.~17(b) we plot $V_x$ for $ 0.43 \leq A \leq 2.5$, 
highlighting the transition from the random   
phase for the two lowest curves at $A = 0.43$ and $A=0.5$ 
to the sliding phase at higher $A$ where all the vortices are moving
and new locking phases for flow along the pinning sites 
occurs. 
This can be seen in the appearance and evolution of the locking steps 
and the reduced fluctuations in $V_{x}$ at higher $A$.

\begin{figure}
\includegraphics[width=3.5in]{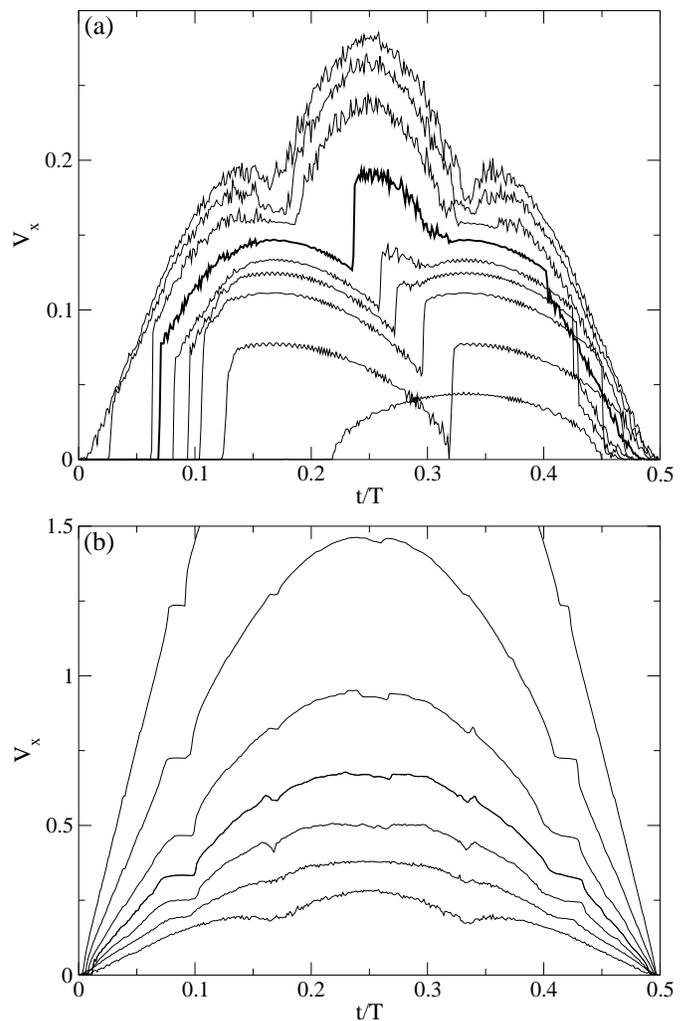}
\caption{
$V_x$ versus time
for different values of the ac amplitude showing the transitions from the
random phases to the high ac drive phases.
(a) From bottom to top, $A = 0.16$, 0.225, 0.3, 0.325, 0.33, 0.35, 0.38, 0.405, 
and  $0.43$.
The $A=0.35$ curve is highlighed with a darker line.
(b) From bottom to top, $A = 0.43$, 0.5, 0.6, 0.75, 1.0, 1.5, and $2.5.$ 
}
\end{figure}

Symmetry locking 
of the type shown in Fig.~10(b)
for high drives 
has been observed previously both for
vortices moving over periodic substrates \cite{Nori}
and in experiments on colloids moving over a square substrate \cite{Grier}. 
In the colloidal studies, the dynamical locking occurred 
in the single particle limit, so the
order-disorder type behavior we observe did not occur. 
We expect that for denser colloidal assemblies 
when a colloidal crystal forms, it should be possible to observe
the order-disorder phenomena we describe here.

\subsection{Commensurability Effects Near the $B/B_{\phi} = 2.0$ State } 

We next examine commensurability effects near the ordered dimer state. 
In Fig.~18(a) we plot $V_{x}$ for $B/B_{\phi} = 1.89$ for a system with the
same parameters as Fig.~1 
where we observed polarization switching of the dimer state.
At the incommensurate filling,
the large interstitial pins contain a mixture of dimers and monomers and the
global orientational ordering of the dimers is lost. 
Those dimers that are present form some local 
ferromagnetic order in one of the three symmetry directions.
Under an applied ac drive, a portion of the dimers
align with the drive; however, the overall disorder in the sample
creates randomness in the energy barriers 
that a dimer must overcome to switch to a new orientation. 
This randomness destroys the sharp global switching of the dimers
shown in Fig.~3 at $B/B_{\phi} = 2.0$ and replaces it with a 
series of smaller avalanche-like switching events. This is illustrated in
Fig.~18(a) where only a portion of the dimers change orientation in any
switching event.
In Fig.~19(a) we show a time trace of the
vortex trajectories for the system in Fig.~18(a) over
several ac drive periods, illustrating that the 
interstitial dimers and monomers remain
pinned but can move within each large interstitial region.
The resulting trajectories are much more random than in the
$B/B_{\phi} = 2.0$ switching case shown in Fig.~2. 
The switching pulses in Fig.~18(a)
do not follow the same pattern for successive periods, indicating
that the avalanche pattern is different from one period to another
and has chaotic-like features.
This type of chaotic avalanche behavior 
is very similar to the Barkhausen noise effects 
found in disordered spin systems \cite{Dahmen}. 
A fuller examination of the Barkhausen or crackling noise effect observed   
in Fig.~18(a)
will be presented in a later work. 

For $B/B_{\phi} = 2.056$, there is a mixture of dimers and trimers in the
large interstitial regions, 
and for $A = 0.075$, a partial depinning of the interstitial vortices occurs
creating the large velocity bursts shown in Fig.~18(b).
The bursts are generally confined to regimes where the 
particles can channel along the six easy-flow directions 
of the pinning lattice, as illustrated in Fig.~19(b). 
For lower values of $A$,
the partial depinning is lost 
and the system has a random switching behavior 
similar to that shown in Fig.~18(a).
For very low ac amplitude, 
even the random switching is lost and the dimers and trimers remain
locked in a fixed orientation.

\begin{figure}
\includegraphics[width=3.5in]{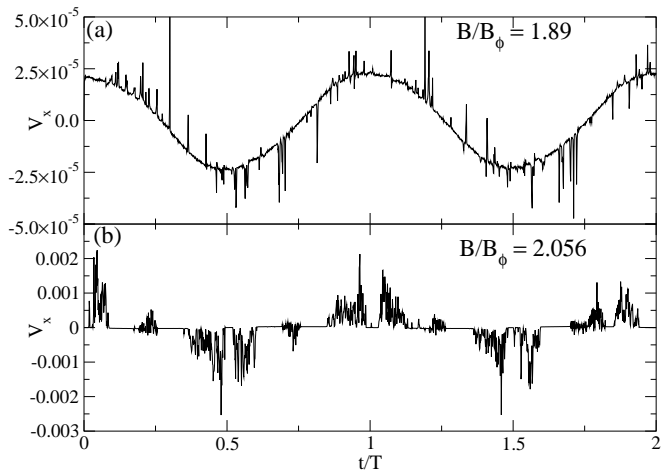}
\caption{
(a) $V_x$ vs time
for a system with the same parameters as in Fig.~2
with $A=0.075$, 
but with $B/B_{\phi} = 1.89$. The vortices all remain pinned 
and the switching from one dimer orientation to another is
random and reminiscent of Barkhausen noise. 
(b) $V_x$ versus time for the same system with $B/B_{\phi} = 2.056$, where
a portion of the vortices begin to slide as the drive is rotated.        
}
\end{figure}

\begin{figure}
\includegraphics[width=3.5in]{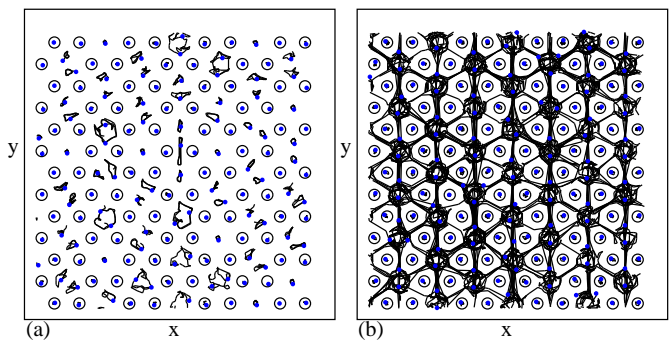}
\caption{
Vortex positions (filled dots), 
pinning site locations (open circles), 
and vortex trajectories (lines) during several periods of motion. 
(a) The system in Fig.~18(a) 
at $B/B_{\phi} = 1.89$. 
The interstitial vortices follow disordered closed orbits. 
(b) The system in Fig.~18(b) at $B/B_\phi=2.056$.
The interstitial vortices depin and move from one interstitial region to
another, while the vortices at the pinning sites remain pinned.
}
\end{figure}

\subsection{ac Effects on the Vortex Trimer State at $B/B_{\phi} = 2.5$} 

As a function of field, higher order vortex molecular crystal states 
form whenever orientationally ordered $n$-mer states occur, 
such as at $B/B_{\phi} = 2.5$, 3.0, or $3.5$. 
These states can show a polarization effect as
a function of the external drive similar to that
observed for $B/B_{\phi} = 2.0$. 
We focus on the case of $B/B_{\phi} = 2.5$, where 
the large interstitial
regions each capture three vortices per pinning site 
and the trimers have a ferromagnetic ordering \cite{Pinning}.
When the external drive is small, the trimers 
remain locked in one orientation without switching.
For larger ac drives, switching of the trimer orientation occurs
that is similar to the switching observed for the dimers. 
For even larger ac drives, a new type of switching effect 
occurs for the trimers that is absent for the
dimer system. 
Within the large interstitial site, the three vortices in the trimer
form a triangle.
When the trimer is aligned along $\theta=\pi/2$ 
as in Fig.~5(c) of Ref.~\cite{Pinning} and the ac drive is also aligned
along $\theta=\pi/2$,
the vortex at the top of the 
triangle experiences an extra force from the lower two vortices
in the triangle and depins into the interstitial site above. 
The same process occurs at every large interstitial site 
and so there is an exchange of one vortex from each site to the site above.
The net result is that the trimers reverse their orientation and are now
pointing along $\theta=3\pi/2$.
The new orientation is more stable against depinning
by the external drive.   
The two vortices at the top of the reoriented trimer produce a barrier for
the motion of the vortex at the bottom of the trimer.  At the same time,
the top two vortices are too wide to hop out of the interstitial site into
the interstitial site above.  This can be regarded as a jamming effect.
As the ac drive cycle continues, the drive rotates into the $\theta=3\pi/2$
direction, and the lower vortex in each trimer can jump into the 
interstitial site below, reorienting the trimer to point along $\theta=\pi/2$.
Thus in this phase, trimer reorientation occurs by the exchange of individual
vortices between neighboring interstitial sites, rather than by
the rotation of the trimer itself.
For higher ac drives, 
a sliding state appears when the vortices move along the symmetry directions of
the pinning lattice.

In Fig.~20 we plot $V_y$ versus time for different values of $A$
in a system at $B/B_{\phi} = 2.5$ with the same parameters as in Fig.~1.
At this filling a set of phases appear which are similar to those found
at $B/B_{\phi} = 2.0$. At very low $A$, the trimers are pinned, while
for larger $A$ the system has sliding states interlaced with pinned states.
A strong channeling effect produces local maxima and minima in $V_y$,
similar to the $B/B_{\phi} = 2.0$ system. 
For $A < 0.15$, the vortices at the pinning sites remain pinned and a 
local minimum in $|V_{y}|$ occurs at $t/T=0.5$.
For $A > 0.15$ this minimum turns into a maximum when the vortices in the
pinning sites begin to depin and generate a random flow phase.
For $A = 0.3$, the transition from a locked flowing phase 
of interstitial vortices to the random phase produces a large jump in $V_y$.
In general, we find that as $B/B_\phi$ increases,
a more complicated set of 
features in the velocity force curves appears.

\begin{figure}
\includegraphics[width=3.5in]{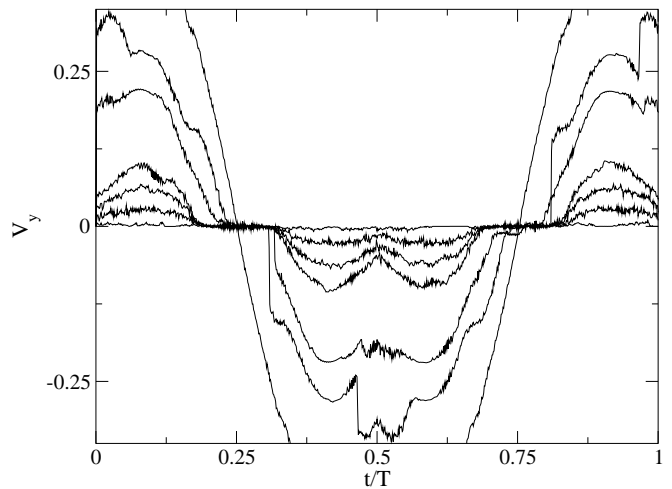}
\caption{
$V_y$ vs time for a system with the same parameters as in Fig.~2 
but with $B/B_{\phi} = 2.5$, where
a trimer state forms. 
From lower left to upper left, 
$A  = 0.075$, 0.1, 0.125, 0.15, 0.25, 0.3, and $0.4$. 
For $A < 0.15$ there is a minimum in $|V_y|$ at $t/T=0.5$, while
for $A \geq 0.25$ a maximum in $|V_y|$ begins to form at $t/T=0$.   
}
\end{figure}

\section{Fixed dc drive in the $y$-direction with an 
ac drive in the $x$-direction }  

In previous work on a honeycomb pinning array at
$B/B_{\phi} = 2.0$, 
we found that
when a dc drive is applied
in the $x$-direction,
the vortices flow along either $\theta=\pi/6$ or 
$\theta=-\pi/6$ depending on the 
original orientation of the dimers in the ground state at $F_d=0$
\cite{Transverse,Transverse2}. 
Once the dimers are flowing under the dc drive, application of an
additional ac drive in the $y$ direction can induce a switching between
flow along $\theta=\pi/6$ and $\theta=-\pi/6$ \cite{Transverse}.
There is a threshold amplitude of the ac drive required to cause the 
switching in the symmetry breaking flow states.
Here we consider the case of a dc drive ${\bf F}^{dc}=F^{dc}{\bf {\hat y}}$
fixed in the $y$ direction with an ac drive 
${\bf F}^{ac} = A\sin(2\pi t/T){\hat {\bf x}}$. 
applied in the $x$ direction.
For
$A = 0$, the dimers flow in one-dimensional 
paths along the $y$ direction and there is no symmetry breaking of the flow,
unlike the symmetry broken flow that occurs for an $x$ direction dc drive.
When an additional ac drive is applied in the $x$ direction, transverse to the
dc drive,
the average value of $V_y$ can be strongly
influenced. 

In Fig.~21 we plot $V_x$ and $V_y$ versus time 
for a system at $B/B_\phi=2.0$ with 
a $y$ direction dc drive of $F^{dc}=0.15$ and an 
$x$ direction ac drive with $A = 0.35$. 
For $0<t/T<0.075$,
the dimer motion is locked along $\theta=\pi/2$.
Although the magnitude of the ac drive is increasing in the positive
$x$ direction during this time period and therefore the overall net driving
force vector is increasing in magnitude, $V_y$ decreases monotonically
with time until a switching event occurs at $t/T=0.075$
when the dimers transition from flowing along $\theta=\pi/2$ to 
flowing along $\theta=\pi/6$.
This switch coincides with a sharp jump in $V_{x}$ to a positive value 
and a cusp feature in $V_{y}$. 
After the switching event, $V_{x}$ and $V_y$ gradually increase with
time until $t/T=0.25$, when they begin to decrease.
For $0.075\leq t/T \leq 0.4$, 
the value of $V_{y}$ is higher than it would be 
under only a dc force and no ac force, indicating that the application of a 
perpendicular ac drive can increase the velocity in the direction of the
dc drive.
For $0.25 < t/T <0.5$, both $V_{x}$ and $V_{y}$ decrease until   
$V_{x} = V_{y} = 0$, indicating that the system is pinned.
The ac drive begins to increase in the negative $x$ direction
above $t/T=0.5$,
and as the dimers reorient to follow the net 
drive vector they realign with $\theta=\pi/2$ and are once again able
to slide along $\theta=\pi/2$ 
for $0.525 < t/T < 0.58$.
At $t/T=0.58$, another switching event occurs and the dimers 
slide along $\theta=11\pi/6$.

\begin{figure}
\includegraphics[width=3.5in]{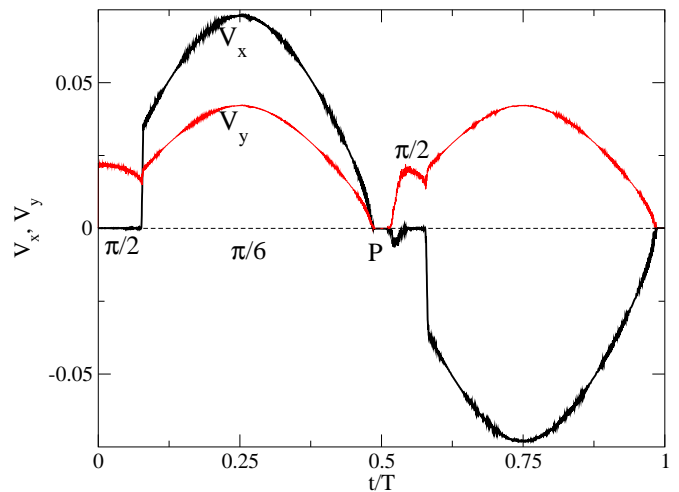}
\caption{
$V_x$ (dark line) and $V_y$ (light line) versus time 
for a system at $B/B_\phi=2.0$ with a $y$ direction dc drive 
of $F^{dc} = 0.15$ and an $x$ direction ac drive 
with $A = 0.35$.
At $t/T=0$, the dimers are moving along $\theta=\pi/2$.
At $t/T=0.075$, a switching transition occurs and the dimers slide along
$\theta=\pi/6$.  Near $t/T=0.5$, a pinned phase (P) appears.
This shows that the application of an ac drive induces a pinned phase 
in a system with a dc drive. 
We note that the ac drive amplitude 
is zero at $t/T=0.5$ and the pinned phase
occurs due to the reorientation of the dimers by the ac drive
into a jammed configuration that is perpendicular to the dc drive.
We call this the jamming transistor effect.    
}
\end{figure}

We term the 
pinning effect in Fig.~21 near $t/T=0.5$ a {\it jamming transistor effect}.
At $t/T=0.5$, 
the ac force is zero in the $x$-direction, so $V_y$ would be expected to take
the value it had at $t/T=0$ when the ac force was also zero. 
Instead, Fig.~21 shows that $V_y=0$ at $t/T=0.5$.
At this point, the ac drive has reoriented the dimers so that they are aligned
along $\theta=0$, instead of along $\theta=\pi/2$ as they were at $t/T=0$.
In previous work on 
honeycomb pinning arrays at $B/B_{\phi} = 2.0$,  we showed 
that if a dc drive is 
applied in the $y$-direction and   
the dimers are aligned perpendicular to the applied driving force, 
an unusually high critical depinning force appears due to the fact that
the dimers are effectively jammed
\cite{Transverse2}. The pinned phase near $t/T=0.5$ 
in Fig.~21 is an example of this
jamming effect.  The dimers act like rigid objects and cannot move through the
$y$-direction channel between adjacent interstitial pinning sites
unless they are aligned along $\theta=\pi/2$.
A key difference between the effect described here and the jamming 
in our previous work is that the jamming in Ref.~\cite{Transverse2}
occurred as an equilibrium transition in the ground state 
when the pinning force was increased. 
In the present case,
the dimers are reoriented dynamically.
In Fig.~22 we schematically illustrate the dynamic jamming effect. 
In Fig.~22(a), a dimer oriented along $\theta=\pi/2$
can move between the pinning sites and has a low depinning threshold.
If the dimer is oriented along $\theta=0$ as in Fig.~22(b), 
parallel to the two pinning sites,
the energy barrier for the dimer to pass between the pinning sites is much
higher and the depinning force is correspondingly higher.

\begin{figure}
\includegraphics[width=2.5in]{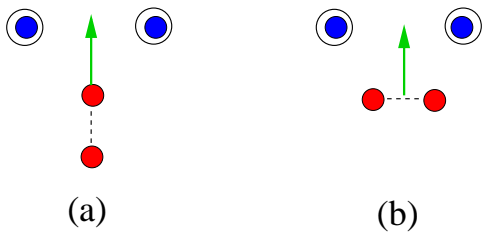}
\caption{
Schematic of the jamming effect. Light circles: dimer vortices in interstitial
site.  Dark circles: pinned vortices.  Open circles: pinning sites.
(a) A dimer oriented along $\theta=\pi/2$
can easily move between two occupied pinning sites and has a low depinning 
threshold. 
(b) A dimer oriented along $\theta=0$ has difficulty moving between the
occupied pinning sites, and the depinning threshold is much higher.
}
\end{figure}

In Fig.~23 we plot $V_{x}$ and $V_{y}$ for a system 
containing only interstitial monomers at $B/B_\phi=1.5$ 
under the same drive configuration as
in Fig.~21, with a dc drive 
applied in the $y$-direction and an ac drive applied in the $x$-direction. 
There is a modulation in $V_y$, with a dip near $t/T=0.12$ at the
point where  $V_x$ rises above zero; however, $V_y$ never drops to zero and
there is no pinned phase.
At $t/T=0.5$, $V_y$ reaches the same value it had at $t/T=0$ at the beginning
of the period when the ac drive component was zero.
This is distinct from the behavior of the dimer system, where $V_y$ dropped
to zero at $t/T=0.5$ due to the reorientation of the dimers.
We find behavior similar to that shown in Fig.~23
over a wide range of ac and dc drive parameter values for
$B/B_\phi=1.5$. 
In general, the application of a force
in the $x$-direction to a system of monomers being driven in the $y$ direction
does not induce a pinned phase or a phase where $V_{y} = 0$. 
The fluctuations in $V_y$ increase when $V_x$ becomes finite because the flow in
this regime is disordered, although the vortices at the pinning sites remain
pinned.
The monomers randomly alternate 
which side of the occupied pinning sites they move around.  

\begin{figure}
\includegraphics[width=3.5in]{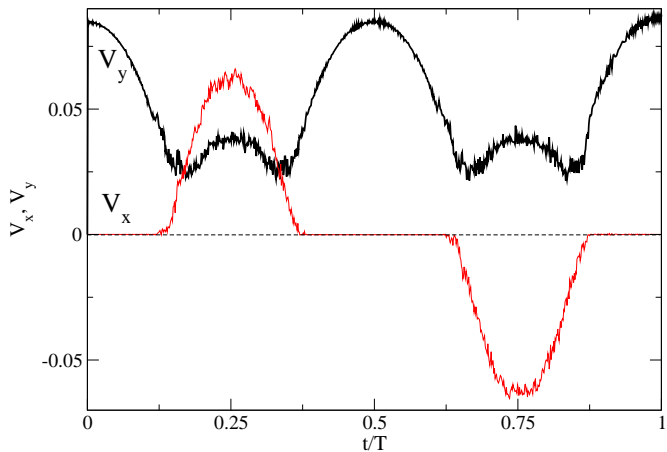}
\caption{
$V_x$ (dark line) and $V_y$ (light line)
for the same system in Fig.~21 with $B/B_{\phi} = 1.5$ 
so that no dimers are present.  
The dc drive is in the $y$ direction with $F^{dc} = 0.425$ 
and a sinusoidal ac drive is applied in the $x$ direction with
$A = 0.35$. Unlike the case of $B/B_{\phi} = 2.0$, 
the velocity responses at $t/T=0$ and $t/T=0.5$ are the same.     
}
\end{figure}

\subsection{Jamming Transistor Effect} 

A system in which a dc velocity component can be switched
on and off with an ac drive
applied perpendicular to the dc drive has similarities to field effect
transistor devices.
The strong cusp in $V_y$ near $t/T=0.075$ in Fig.~21
and the corresponding sudden jump in $V_{x}$ suggest 
that the rapid switching time could be used to create
semiconductor type devices similar to the 
transverse field effect depinning threshold modulation 
in charge density wave systems \cite{Toner,Dohmen}.    
Since the external ac field can control the dimer orientation and since the
depinning threshold depends on the orientation of the dimers, 
the system should exhibit strong memory effects of its original 
configuration.
For example, different settings of the initial ac and dc drive values
would orient the dimers in a particular direction and
the pinning threshold would depend on the history of the           
preparation of the dimer states.  
These field effect type properties could be 
useful for making vortex-based devices 
or could be used as models for constructing 
similar systems on smaller scales using 
ions or Wigner crystals.  

We showed in Fig.~23 that monomers do not exhibit a jamming effect. 
In general, the jamming effect occurs when $B/B_\phi$ is
close to $B/B_{\phi} = 2.0$.
As long as there are enough interstitial dimers to 
percolate throughout the entire sample, the dimers can induce a jamming
effect which has a side effect of pinning 
any monomers or trimers that are present.
Even at $B/B_{\phi} = 2.0$, there
is a threshold value of $F^{dc}$ above which the jamming 
transition no longer occurs, as we describe in Section IV C.
For fields away from $B/B_{\phi} = 2.0$, this dc threshold force is reduced. 
For $B/B_{\phi} < 1.75$, a complete 
jamming effect is lost; however, 
portions of the system populated with dimers can still jam. For 
$B/B_\phi<2.0$ and low enough
$F^{dc}$, the monomers are simply pinned at the interstitial sites
and jamming of the dimers has no effect on the monomer pinning.

\subsection{Varied ac Amplitude and Fixed dc Amplitude} 

We next examine the effect of varied ac amplitude for a system with 
fixed $F^{dc} = 0.1$ 
applied in the $y$ direction at $B/B_{\phi} = 2.0$. 
If the ac amplitude is too small,
the vortex motion remains locked along $\theta=\pi/2$ in the $y$-direction. 
In Fig.~24(a) we plot $V_{x}$ and $V_{y}$ for $A = 0.16$.  
At this ac drive value, the
jump in $V_{x}$ is close to its maximum amplitude,
and the corresponding cusp feature in $V_{y}$ is reduced
in size. The pinned phase
near $t/T=0.5$ is also more extended than in the sample illustrated
in Fig.~21. In Fig.~24(b) we show $V_x$ and $V_y$ in the same system at
$A = 0.15$, which is just below the
ac threshold required to unlock the motion from the $y$-direction, 
$\theta=\pi/2$. 
In this case $V_{x} = 0$ throughout the cycle while $V_y$ 
has a periodic modulation but is always finite.
Even though the vortices are not translating in the $x$-direction, 
the application of 
an ac force in the $x$-direction causes the vortices flowing along the $y$
direction to shift toward one side of the $\theta=\pi/2$ 
flow channel.   This increases the strength of the interaction
between the moving vortices and the pinned vortices 
and increases the effective drag on the moving vortices, reducing $V_y$
whenever the absolute value of the ac drive is 
maximum in the
positive or negative $x$-direction.
In Fig.~25 we show a series of $V_{y}$ versus time
curves for increasing ac drive amplitude with $A = 0.35$, 0.45, 0.55, and 
$0.85.$ 
As $A$ increases, the transition from $\theta=\pi/2$ to $\theta=\pi/6$ motion
occurs at earlier $t/T$.
For $ A > 0.35$, there is a large jump into a fluctuating phase
centered near $t/T=0.25$, corresponding to the depinning of a portion of the
pinned vortices into a random flow regime.
The system can then reorganize to the $\theta=\pi/6$ flow phase 
at a later point in the period when the force from the ac 
drive goes to zero and the net driving force is no longer large enough to
depin vortices from the pinning sites.
The vortices enter a pinned phase at $t/T=0.5$.

\begin{figure}
\includegraphics[width=3.5in]{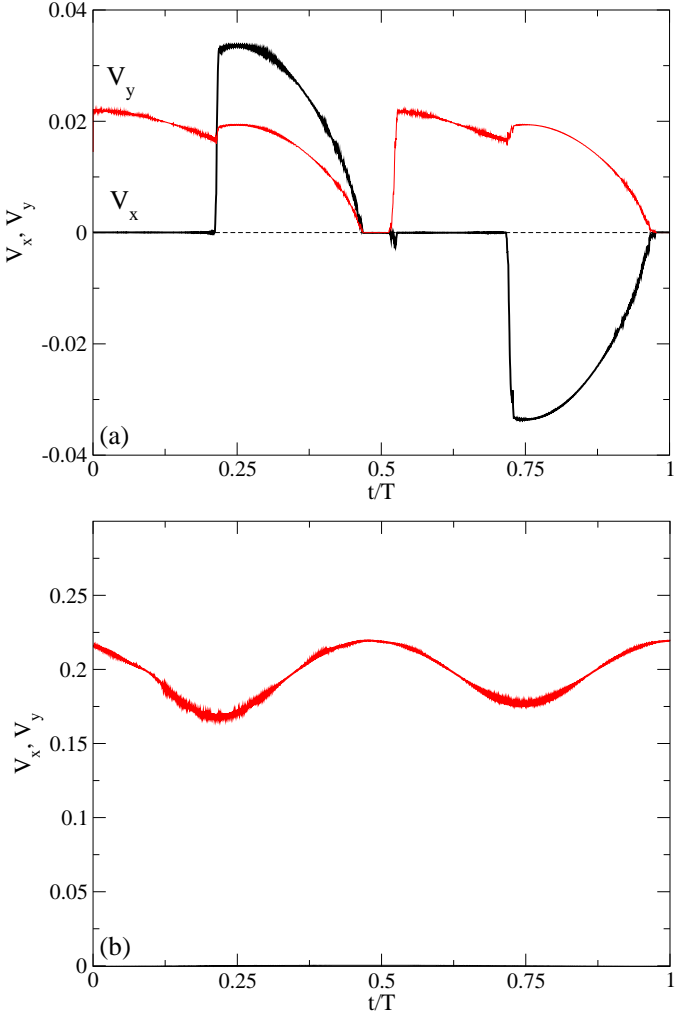}
\caption{
(a) $V_x$ (dark line) and $V_y$ (light line) vs time
for a system with a $y$ direction dc force of 
$F^{dc} =  0.15$ and an $x$ direction ac force
with $A = 0.16$.
Here the jamming transition
appears along with a strong switching effect in $V_{x}$.
(b) The same system at $A = 0.15$, where the ac 
amplitude is too small to induce a switching effect
and $V_x=0$.     
}
\end{figure}

\begin{figure}
\includegraphics[width=3.5in]{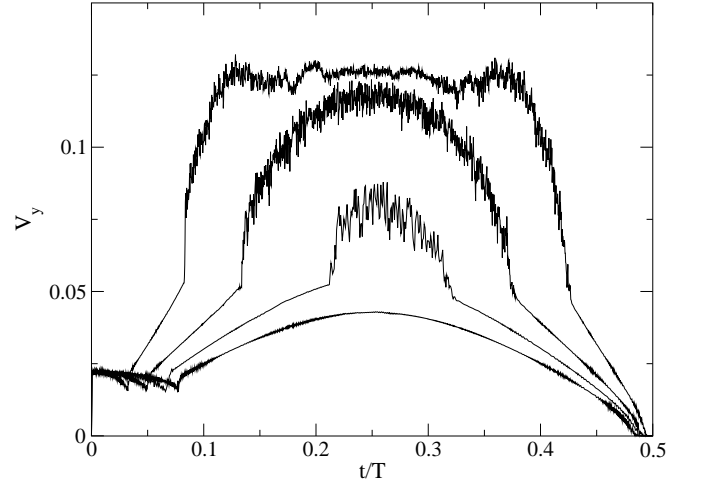}
\caption{$V_{y}$ vs time for the same system in Fig.~21 
with $F^{dc}=0.15$ at 
$A = 0.35$, 0.45, 0.55, and $0.85$, from bottom to top.    
The large jump for $A > 0.35$ near $t/T=0.25$ 
is the onset of the random phase where vortices
at the pinning states start to depin.  
}
\end{figure}

In Fig.~26  we show the dynamical phase diagram of time versus ac
amplitude $A$ during half of the ac drive
period with fixed $F^{dc}=0.15$.
At small $t/T$ when the ac force is first increasing from zero, the system
always starts with the dimers flowing along $\theta=\pi/2$.
For decreasing $A$, the temporal extent of the $\theta=\pi/2$ 
region grows until for $A \leq 0.15$ 
the system is always locked to $\theta=\pi/2$,
as shown in Fig.~24(b) for $A = 0.15$.    
For $0.15 < A < 0.45$, the system passes through $\theta=\pi/2$ and
$\theta=\pi/6$ locked flow phases 
and a jammed or pinned phase in half a period.
The random flow phase first appears for $A\geq 0.45$.
In the random flow phase, vortices at the pinning sites depin when
the net driving force from the combined dc and ac drives is maximum, 
which occurs at $t/T=0.25$.
The jammed phase which is centered at $t/T=0.5$
becomes narrower but still persists for increasing $A$.
 For $A > 1.25$ (not shown) there are new types of 
moving phases that arise within the random phase region 
when all the vortices depin and 
start sliding along the pinning sites, producing the
order-disorder transitions studied earlier. 

\begin{figure}
\includegraphics[width=3.5in]{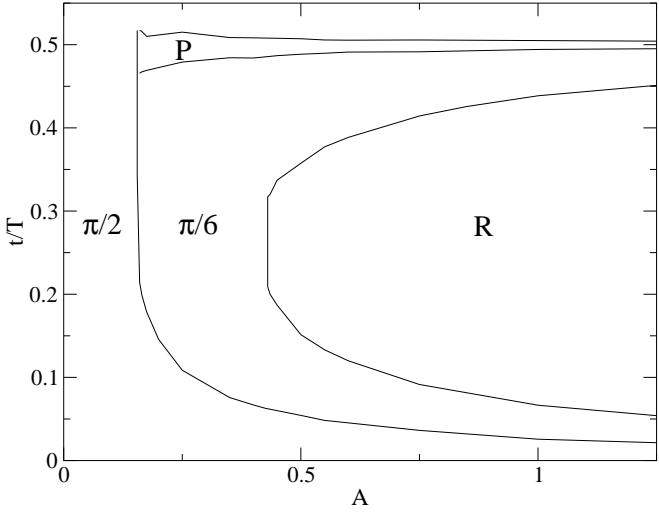}
\caption{
The dynamical phase diagram for time $t/T$ vs ac amplitude $A$
for the system in Fig.~21 with a fixed dc drive 
of $F^{dc}=0.15$ in the $y$-direction and
an ac drive applied in the $x$-direction. 
In the $\theta=\pi/2$ phase the dimers flow along the
$y$-direction, while for higher $A$ they flow along $\theta=\pi/6$. 
The pinned phase is marked P while R is the random phase 
where the vortices start to depin from the pinning sites.    
}
\end{figure}

\subsection{Constant ac Amplitude and Varied dc Amplitude} 

We next consider the case where the 
$x$ direction ac drive amplitude is fixed at $A=0.2$
and the $y$ direction dc drive amplitude is varied 
at $B/B_{\phi} = 2.0$.
At $F^{dc} = 0,$ the behavior of the system resembles 
that observed in earlier studies with only a dc drive in the
$x$-direction \cite{Transverse,Transverse2}. In this limit the ac drive 
applied along the $x$ direction induces flow along $\theta=\pi/6$ or
$\theta=11\pi/6$, with the flow reversing 
during the second half of the ac cycle to $\theta=7\pi/6$ or 
$\theta=5\pi/6$,
respectively.
In Fig.~27(a) we plot  $V_{x}$ and $V_{y}$ for a system with
$F^{dc} = 0$ and $A = 0.2$. 
Here the vortices are initially pinned and then undergo a transition
to the moving $\theta=11\pi/6$ phase.
In this case, the dimers were oriented along $\theta=11\pi/6$ in 
the initial pinned ground state.  
The amplitude of $V_{x}$ is larger than $V_{y}$ since 
$|V_{x}| = |A\cos(11\pi/6)|$ and 
$|V_{y}| = |A\sin(11\pi/6)|$. 
When the ac drive reverses, 
the dimers flow along $\theta=5\pi/6$.
The symmetry breaking of the dimers in the ground state fixes 
the flow direction of the dimers for all later times. 
If we consider an incommensurate case 
where there are additional disordered phases, the system dynamically
organizes into one of the broken symmetry phases. 
In this case, when the ac drive reverses sign,
the system passes though the disordered
phase which destroys the memory of the dynamically broken symmetry phase,
and the vortices have an equal chance of dynamically breaking symmetry
in either of the two possible directions.     

\begin{figure}
\includegraphics[width=3.5in]{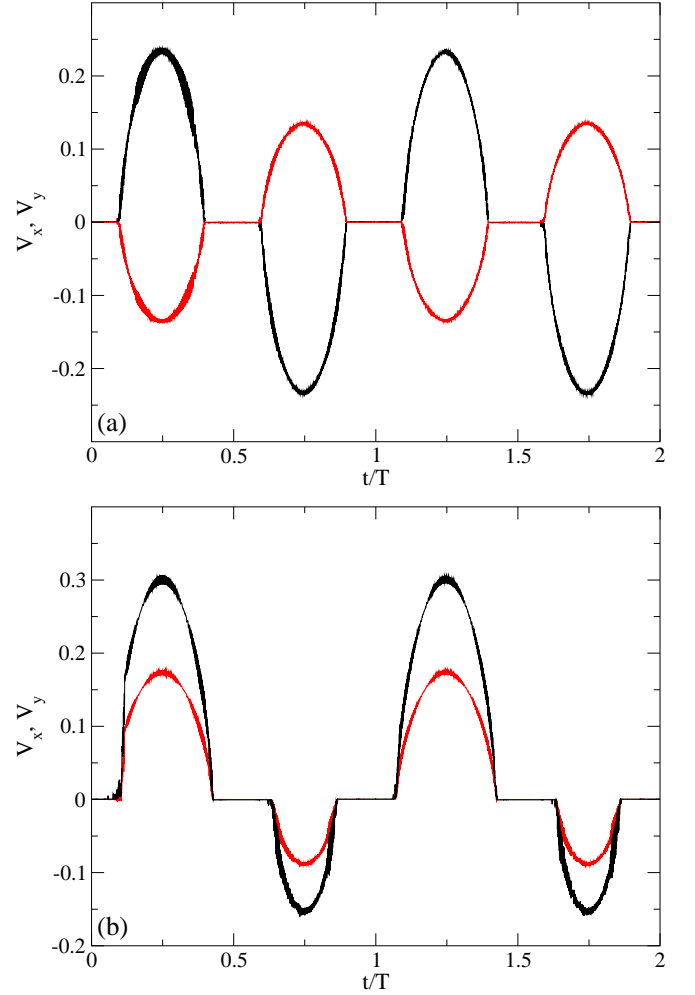}
\caption{
$V_{x}$ (dark line) and $V_{y}$ (light line) vs time for 
a system at $B/B_{\phi} = 2.0$
with a $y$ direction dc drive and an $x$ direction sinusoidal ac drive
with $A = 0.2.$ 
(a) $F^{dc} = 0$. Here the vortex
motion is locked along $\theta=11\pi/6$ or $5\pi/6$ during a portion of the
first and second halves of each ac drive cycle.  (b) $F^{dc}=0.05$. Here,
the dc drive is not strong enough to lock the motion 
to only the positive $y$-direction, and the direction of
motion is determined by the symmetry breaking of the 
dimers in the ground state.       
}
\end{figure}

At $B/B_{\phi} = 2.0$,  if a small but finite dc drive 
$F^{dc}$ is applied in the
$y$ direction, it is unable to
break the symmetry of the ground state, 
and the ground state dimer orientation
determines the resulting dynamics, 
as illustrated in Fig.~27(b) 
for a system with $F^{dc} = 0.05$. 
Here the vortices are initially pinned and then
depin to flow along $\theta=\pi/6$,
reflecting the orientation of the dimers in the
pinned state. 
During the second half of the drive period, the dimers flow along
$7\pi/6$
and $|V_x|$ and $|V_y|$ are both lower than they were during the first
half of the drive period.
This is due to the biasing force of the dc drive component 
which is applied in the positive
$y$ direction.      
When $F^{dc}$ is sufficiently strong, the negative component of the $y$ velocity
during the second half of the drive period
is lost. 

In Fig.~28 we plot $V_{y}$ versus time for $A = 0.2$ 
and $F^{dc} = 0.05$, 0.125, 0.2, 0.225,
and $0.275$. 
For $F^{dc} > 0.125$, $V_{y}$ is always positive.  
For $F^{dc}>0.1$, the vortices initially 
move along $\theta=\pi/2$ and as $F^{dc}$ increases, the
magnitude of $V_{y}$ increases, as does 
the magnitude of the jump from $\theta=\pi/6$ back to $\theta=\pi/2$
near $t/T=0.5$.
At $F^{dc} = 0.275$, the $\theta=\pi/6$ phase is replaced with 
a randomly fluctuating phase when the net force from the combined dc and
ac drives
is strong enough to depin vortices from the pinning sites, 
similar to the behavior shown earlier.
For $F^{dc} < 0.2$, $V_{x}$ (not shown) steadily rises with $t/T$ and 
reaches a maximum value near
$t/T=0.25$ which saturates for $F^{dc} > 0.2$. 
The saturation occurs since the ac component
in the
$x$-direction is fixed. 
The onset of the random phase also produces increased fluctuations
in $V_{x}$ at $F^{dc} = 0.275$. 

\begin{figure}
\includegraphics[width=3.5in]{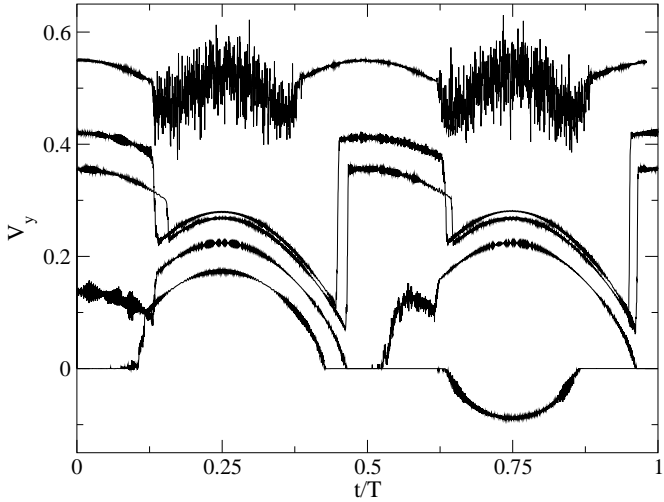}
\caption{
$V_{y}$ vs time
for the same system in Fig.~27 with an
$x$ direction ac drive of $A = 0.2$ 
and a $y$ direction dc force.
From bottom to top, $F^{dc} = 0.05$, 0.125, 0.2, 0.225, and 0.275. 
The onset of the random 
phase where vortices depin from the pinning sites appears as a jump
into a strongly fluctuating phase  for 
the $F^{dc} = 0.275$ curve. 
}
\end{figure}

In Fig.~29(a) we show the continued evolution of $V_{y}$ from Fig.~28(a) for 
$F^{dc} = 0.33$, 0.41, 0.45, 0.5, 0.55, and $0.6$.
The $\theta=\pi/2$ sliding phase diminishes in size with increasing
$F^{dc}$ and vanishes completely
for $F^{dc} > 0.475$ while
the random flow phases grow in extent. 
Within the random flow phases for $F_{dc} > 0.475$, 
there is a modulation of $V_y$
with dips near $t/T=0$, $t/T=0.5$, and $t/T=1$, 
along with smaller modulations near $t/T=0.25$ and $t/T=0.75$. 
It is interesting to note that the
ratio of $V_y$ at $t/T=0.25$ to $V_y$ at $t/T=0.5$ reaches a maximum at
$F^{dc} = 0.45$. 
Here $V_y$ is strongly reduced 
when the system enters the $\theta=\pi/2$ phase because the vortices that
depinned during the random flow phase repin in the $\theta=\pi/2$ phase
and no longer contribute to $V_y$.
In Fig.~29(b) we show the corresponding $V_{x}$ curves which indicate
that the $\theta=\pi/2$ locked flow phase
disappears for $F^{dc} > 0.475$.  
  
\begin{figure}
\includegraphics[width=3.5in]{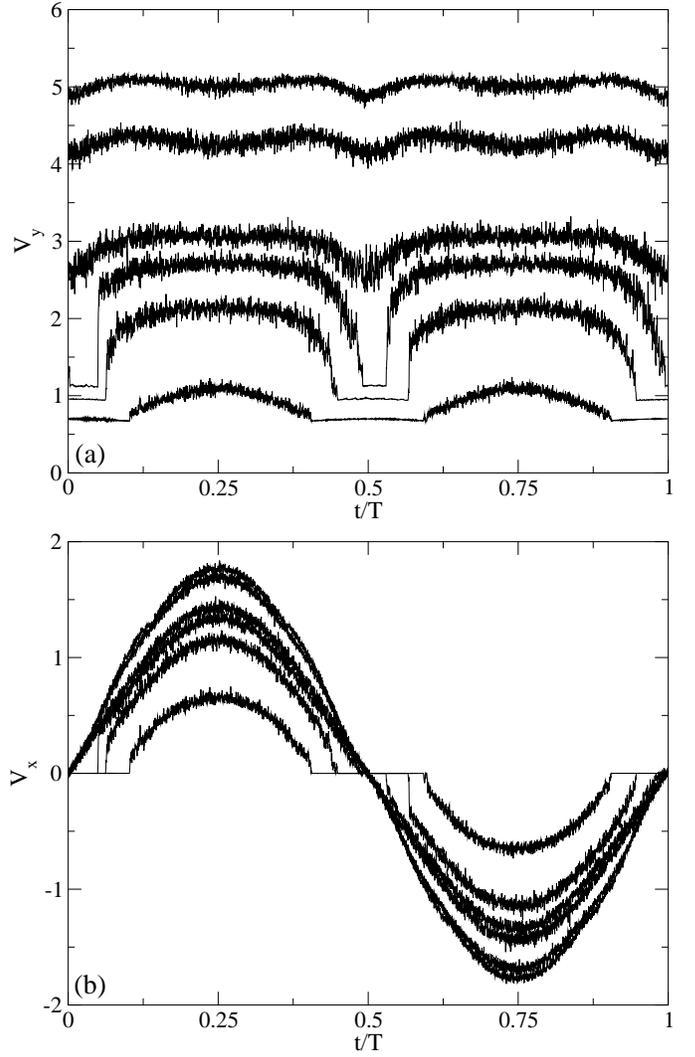}
\caption{
(a) $V_y$ vs time for the system in Fig.~28 at
$A = 0.2$ and $F^{dc} = 0.33$, 0.41, 0.45, 0.5, $0.55$, and 0.6 (from bottom
to top).
For $F^{dc} > 0.45$, the sliding phases $\theta=\pi/2$ 
at $t/T=0$ and $t/T=0.5$ are lost. 
(b) The corresponding $V_{x}$ vs time at $F^{dc}=0.33$, 0.41, 0.45, 0.5,
0.55, and 0.6 (from bottom to top at $t/T=0$).   
}
\end{figure}

In Fig.~30 we plot the dynamic phase diagram for 
$F^{dc}$ versus time at
fixed $A = 0.2$ for the system in Figs.~28 and Fig.~29 
at $B/B_\phi=2.0$.  
For $t/T<0.1$ and $F^{dc} < 0.075$, the vortices are initially 
in the pinned phase.  As $t/T$ increases and the ac force becomes larger,
the vortices organize into a 
pinned symmetry broken state for $F^{dc}<0.075$ or
depin into the $\theta=\pi/6$ sliding state for $0.075\leq F^{dc}<0.475$.
The initial pinned state is lost for $F^{dc} > 0.1$  
when the dc drive is strong enough 
to cause the dimers to slide along $\theta=\pi/2$.
For $F^{dc} > 0.475$, the dc drive combined with the motion of 
the interstitial vortices is  strong enough to depin 
the vortices at the pinning sites and the system enters
the random phase even when $F^{ac} = 0$ at $t/T=0$.
For $0.1\leq F^{dc} \leq 0.25$, the system switches from 
$\theta=\pi/2$ to $\theta=\pi/6$
dimer flow, while the $\theta=\pi/6$ phase is lost for $F^{dc} \geq 0.25$ 
and the system enters
the random flow state. 
The second pinned state near $t/T=0.5$ disappears for $F^{dc} > 0.2$.
Due to the polarization of the dimers, this is higher than 
the drive at which the pinned state disappears for
$t/T=0$.
For higher dc amplitudes than those shown in Fig.~30, 
all of the vortices depin
and slide along certain directions of the pinning lattice, forming
the order-disorder transitions studied earlier.  

\begin{figure}
\includegraphics[width=3.5in]{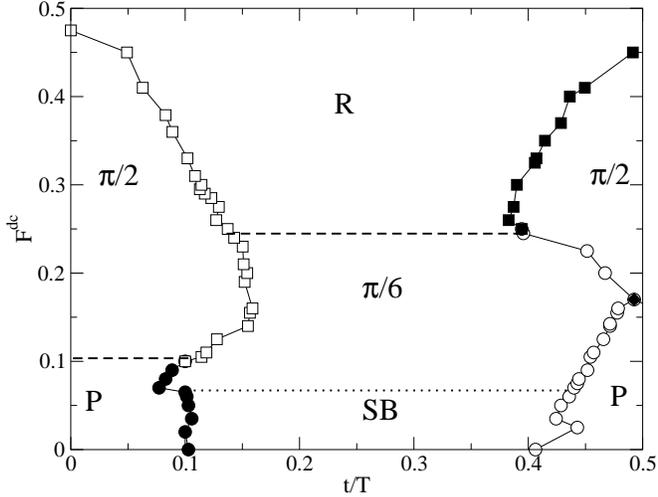}
\caption{
Dynamic phase diagram $F^{dc}$ versus time
for the same system in Figs.~28 and Fig.~29 at $B/B_\phi=2.0$ under
an $x$ direction ac drive with $A  = 0.2$ and 
a $y$ direction dc drive.
P: pinned phase; SB: symmetry breaking phase where the symmetry is broken in
the pinned dimer configuration; R: random phase where vortices at the
pinning sites become depinned.  There are two sliding phases along 
$\theta=\pi/2$ and $\theta=\pi/6$.
}
\end{figure}

\subsection{Commensurability Effects}

We next consider the effects
of varied $B/B_{\phi}$. 
Figure~31 shows $V_{y}$ versus 
time for $B/B_{\phi} = 2.44$, 3.0, 3.4, and 4.72 
for a system with fixed 
$F^{dc} = 0.35$ applied in the $y$-direction 
and an ac force with $A=0.2$ applied in the $x$-direction.
For clarity, 
the velocities are normalized by the value of $V_y$ at $B/B_\phi = 2$. 
For $B/B_{\phi} = 2.44$, the system is in the random phase
throughout the driving period; however, a prominent 
channeling effect persists as shown by the minima at $t/T=0.5$
and $t/T=1.0$ as well as the smaller shoulders at 
$t/T=0.25$ and $t/T=0.75$. 
For $B/B_{\phi} = 3.0$,
the system forms several  new ordered flow phases
visible as regions of reduced fluctuations in $V_y$. 
The ordered interstitial flow phase ends at $t/T=0.04$ with a
transition into a random flow regime, as
shown by the drop in $V_{y}$. 
The same ordered interstitial flow state 
reappears just before and after $t/T=0.5$, as indicated by the sharp cusps
in $V_y$, and also occurs 
for $B/B_{\phi} = 3.4$.
In Fig.~32 we illustrate the vortex trajectories in the ordered 
interstitial flow phase for 
$B/B_\phi = 3.4$.
The moving interstitial vortices channel in periodic winding paths 
at an angle between the
occupied pinning sites. 
At this particular filling, a small portion of the interstitial
vortices are trapped behind the occupied pinning sites. 
The ordered interstitial 
flow phase at $B/B_\phi=3.0$ is nearly identical except that the
immobile interstitial vortices are not present. 

\begin{figure}
\includegraphics[width=3.5in]{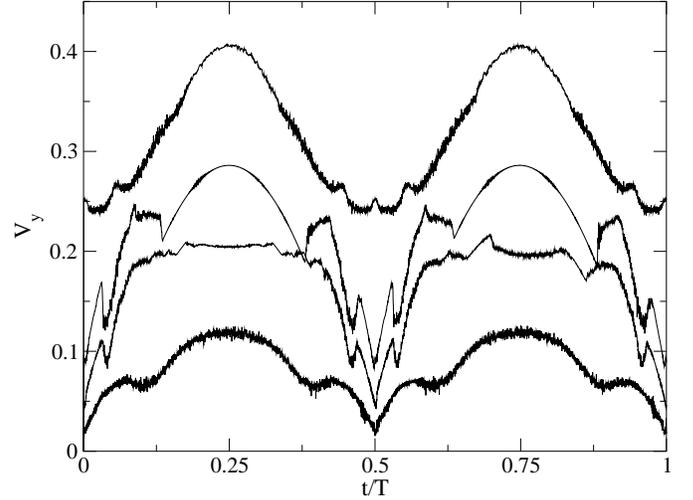}
\caption{
$V_{y}$ vs time at varied $B/B_\phi$ 
with a $y$ direction dc drive of
$F^{dc} = 0.35$ 
and an $x$ direction ac drive with $A = 0.2$.
For clarity, the velocities are normalized by the value of $V_y$ 
at $B/B_{\phi} = 2.0$.
From top to bottom, $B/B_{\phi} = 2.44$, 3.0, 3.4, and $4.72$.
For $B/B_{\phi} = 3.4$, a complex cusp structure forms 
and two different types of ordered phases appear.
}
\end{figure}

\begin{figure}
\includegraphics[width=3.5in]{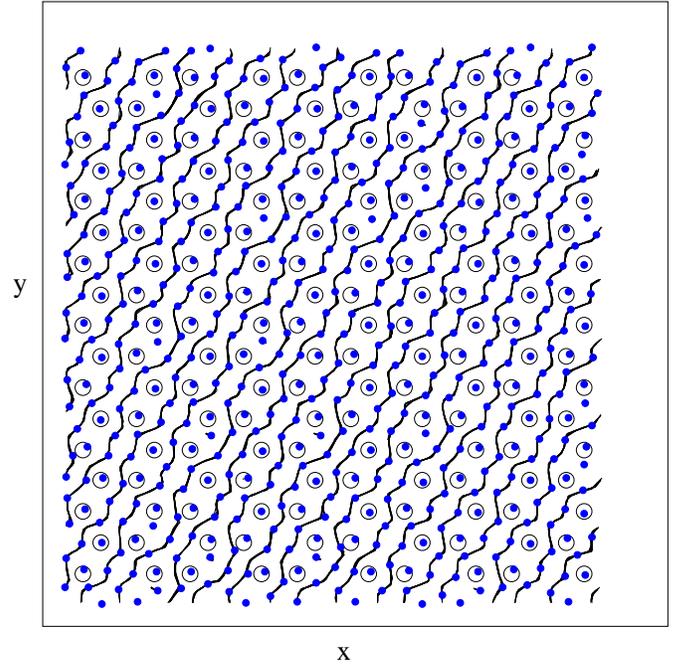}
\caption{
Vortex positions (filled dots), pinning site locations (open circles),
and vortex trajectories (lines) in the ordered 
interstitial flow phase at $t/T=0.025$
for the $B/B_{\phi} = 3.4$ system in Fig.~31. 
Here the moving interstitial vortices form ordered channels 
between the pinning sites.
A small portion of the interstitial vortices 
are pinned behind occupied pinning sites.
This same type of ordered interstitial flow phase occurs at $B/B_\phi=3.0$.
}
\end{figure}

In Fig.~31, as the ac force increases for $B/B_{\phi} = 3.0$ 
the vortices at the pinning sites depin into a floating 
triangular structure and the flow becomes random at $t/T=0.1$.
Near $t/T=0.15$, the fluctuations in $V_y$ are reduced when a portion
of the vortices lock into a channeling motion 
along the pinning rows. 
For $B/B_{\phi} = 3.4$, the same floating triangular
lattice appears; however,
the transition from the random phase to the 
channeling phase centered at 
$t/T=0.25$ is much sharper. 
The plot of $P_6$ versus time for the $B/B_\phi=3.4$ system in Fig.~33
shows the transition to the floating triangular state, and also indicates
that the vortex lattice regains some order during the ordered 
interstitial flow phases.
The  transitions into the disordered regimes appear as drops in $P_{6}$.  

\begin{figure}
\includegraphics[width=3.5in]{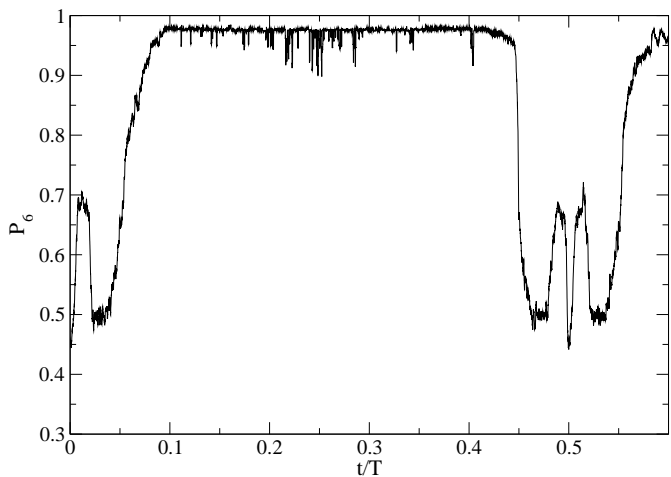}
\caption{
The fraction of six-fold coordinated vortices $P_{6}$ vs time for
the $B/B_{\phi} = 3.4$ system from Fig.~31. 
The cusp structures in $V_y$ in 
Fig.~31 correlate with the features in $P_{6}$. In the channeling
phase centered at $t/T=0.25$, the vortices form a mostly triangular  
lattice. 
}
\end{figure}

For $B/B_{\phi} = 4.72$ in Fig.~31, random flow phases
dominate the behavior; however, several cusp features still appear
in $V_y$.
The vortex lattice is also mostly triangular near 
$t/T=0.25$.
A new feature at this field is the transformation of the minima in
$V_y$ at $t/T=0.5$ into a small maximum.
These results show that there is a complex 
variety of phases at the different fields which have clear
transport and ordering signatures.

\section{Discussion} 
In this work we considered varied ac drive amplitude 
but kept the ac drive frequency fixed. 
Our results should
be robust in the low frequency regime, and 
for frequencies lower than the frequency considered here, 
the results do not change.
At much higher frequencies,
the system does not have time to respond to the ac drive
and the switching events will be lost.
In our previous work, we found that for a frequency about 
10 to 100 times higher
than the frequency considered here, the system stops responding 
to the ac drive \cite{Transverse}.       
In Fig.~34 we show an example of this effect for a system with a dc 
drive applied in the $y$-direction
and an ac drive applied in the $x$-direction, 
with the same parameters as the system in Fig.~23 but at a higher
frequency. For
low frequency, the system shows a switching effect from 
$\theta=\pi/2$ flow to $\theta=\pi/6$ flow.
In Fig.~34 the frequency is $10^3$ times higher and the 
vortex trajectories show oscillations; however, 
the net motion remains locked to $\theta=\pi/2$.

\begin{figure}
\includegraphics[width=3.5in]{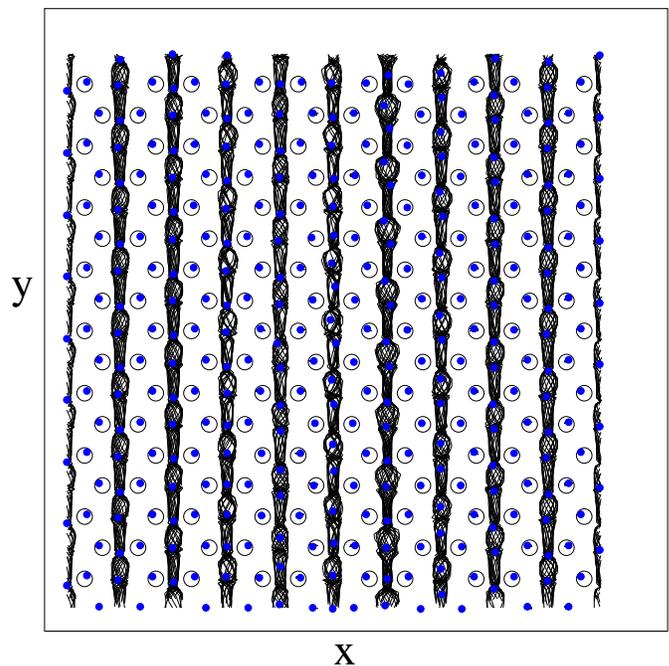}
\caption{
Vortex positions (filled dots),
pinning site locations (open circles), and vortex trajectories (lines)
during several drive periods for the same system in Fig.~23 with a 
$y$ direction dc drive of $F^{dc}=0.425$
and an $x$ direction ac drive with $A=0.35$.
In Fig.~23 the frequency was $10^3$ times lower and the
system exhibited a switching from $\theta=\pi/2$ flow 
to $\theta=\pi/6$ flow.  Here, the
ac frequency is higher than the ability of the vortices to respond,
and the motion stays locked in the $\theta=\pi/2$ direction.
}
\end{figure}

Our previous work with this system showed that temperature effects 
can change the phase boundaries slightly, but that the ordered dynamical 
phases persist until the melting transition of the interstitial vortices 
is reached \cite{Transverse,Transverse2}. 
In the present work, we focused on the limit where the pinning is strong 
enough that there is a regime in which
the vortices at the pinning sites can remain pinned 
while the interstitial vortices flow. 
As $F_{p}$ decreases, many of the dynamical phases will disappear as 
the dimers lose their independent effective orientational degree of 
freedom.  For low enough $F_p$, additional
transitions can occur in the vortex lattice 
where a portion of the formerly pinned vortices are no longer located
at the pinning sites, producing a triangular structure 
\cite{Peeters1,Pinning,Misko3}. 

\section{Summary} 
We have shown that in honeycomb pinning arrays, 
vortex molecular crystalline states 
which have effective orientational degrees of freedom  
can have their orientation controlled by an external ac drive.
We specifically focused on vortex dimers at the second matching
field which have three different possible orientations.
When an ac drive is applied in both the $x$ and $y$ directions, the     
orientation of the dimers can be made to follow 
the external drive while the vortices remain pinned in the large 
interstitial site.  We term this a polarization effect. 
In this regime
the dimers 
rotate by a combination of continuous motion and 
a switching type behavior
that appears as large velocity jumps 
when the dimers switch to the different symmetry angles of the
pinning array. 
For very weak ac drive amplitudes, the dimers remain locked 
in their initial orientation, showing that
there is a threshold force needed to induce the pinned switching behavior.
For large ac drive amplitudes, the dimers
depin and flow between the pinned vortices
along certain symmetry directions of the pinning lattice, 
producing distinctive features in the $x$ and $y$ components of the velocity 
during each ac period. For even larger 
ac drive amplitudes, all the vortices depin and the system 
exhibits a series of order-disorder transitions 
and symmetry locking effects as a function of ac drive.  
We also consider the case of a fixed dc drive applied in one direction 
and an ac drive applied in the perpendicular direction. 
Here,
the application of the ac drive can cause the moving dimers to 
orient in such a way that they become effectively jammed. 
We call this effect a jamming transistor and 
show that it occurs due to the fact that the dimers 
can slide more easily if they are oriented 
along the symmetry axis of the pinning rows. 
The ac drive can also induce a  
pronounced dynamical switching that can be seen in both velocity components. 
Many of the switching features 
may be useful for creating new types of vortex based 
devices or fluxtronics, and
can serve as examples for other systems in which
molecular crystal type states can form, such 
as ions or Wigner crystals on periodic substrates, 
colloidal molecular crystals, or vortices in Bose-Einstein condensates.   

\section{Acknowledgments}
We thank M. Hastings for useful discussions. 
This work was carried out under the NNSA of the U.S. DOE at LANL under
Contract No. DE-AC52-06NA25396.


\begin{thebibliography}{99}

\bibitem{Brandt}
E.H.~Brandt, J.~Low.~Temp.~Phys.~{\bf 53}, 41, 71 (1983);
H.J.~Jensen, A.~Brass, and A.J.~Berlinsky, Phys.~Rev.~Lett.~{\bf 60}, 
1676 (1988);
H.J.~Jensen, A.~Brass, Y.~Brechet, and A.J.~Berlinsky, 
Phys.~Rev.~B {\bf 38}, 9235 (1988). 

\bibitem{Koshelev}
A.E. Koshelev and V.M. Vinokur, Phys.~Rev.~Lett.~{\bf 73}, 3580 (1994).  

\bibitem{Giamarchi}
P. Le Doussal and T.~Giamarchi, Phys.~Rev.~B {\bf 57}, 11 356 (1998). 

\bibitem{Balents}
L.~Balents, M.C.~Marchetti, and L.~Radzihovsky, 
Phys.~Rev.~Lett.~{\bf 78}, 751 (1997);
Phys.~Rev.~B {\bf 57}, 7705 (1998). 

\bibitem{Moon}
K.~Moon, R.T.~Scalettar, and G.T.~Zim{\' a}nyi, 
Phys.~Rev.~Lett.~{\bf 77}, 2778 (1996);
S.~Spencer and H.J.~Jensen, Phys.~Rev.~B {\bf 55}, 8473 (1997); 
C.J.~Olson, C.~Reichhardt, and F.~Nori, 
Phys.~Rev.~Lett.~{\bf 81}, 3757 (1998); 
A.B.~Kolton, D.~Dom{\' i}nguez, and N.~Gr{\o}nbech-Jensen, 
{\it ibid.}~{\bf 83}, 3061 (1999).  

\bibitem{Higgins}
S.~Bhattacharya and M.J.~Higgins, Phys.~Rev.~Lett.~{\bf 70}, 2617 (1993);
A.C.~Marley, M.J.~Higgins, and S.~Bhattacharya, 
{\it ibid.}~{\bf 74}, 3029 (1995). 

\bibitem{Pardo}
F.~Pardo, F.~de la Cruz, P.L.~Gammel, E.~Bucher, and D.J.~Bishop,
Nature (London) {\bf 396}, 348 (1998); 
A.M.~Troyanosvki, J.~Aarts, and P.H.~Kes, 
{\it ibid.}
{\bf 399}, 665 (1999).  

\bibitem{Soret}
E.~Olive and J.C.~Soret, Phys.~Rev.~B {\bf 77}, 144514 (2008).  

\bibitem{Mangan}
N.~Mangan, C.~Reichhardt, and C.J.~Olson Reichhardt, 
Phys.~Rev.~Lett.~{\bf 100}, 187002 (2008). 

\bibitem{Reichhardt}
C. Reichhardt, C.J.~Olson, and F.~Nori, 
Phys.~Rev.~Lett.~{\bf 78}, 2648 (1997);   
Phys.~Rev.~B {\bf 58}, 6534 (1998). 

\bibitem{Zimanyi}
C.~Reichhardt, G.T.~Zim{\' a}nyi, and N. Gr{\o}nbech-Jensen, 
Phys.~Rev.~B {\bf 64}, 014501 (2001). 

\bibitem{Carneiro}
G.~Carneiro, Phys.~Rev.~B {\bf 62}, R14 661 (2000);  
{\bf 66}, 054523 (2002). 

\bibitem{Zhu}
B.Y.~Zhu, L.~Van Look, V.V.~Moshchalkov, B.R.~Zhao, and Z.X.~Zhao,
Phys.~Rev.~B {\bf 64}, 012504 (2001).  

\bibitem{Harada}
K.~Harada, O.~Kamimura, H.~Kasai, T.~Matsuda, A.~Tonomura, 
and V.V.~Moshchalkov,
Science {\bf 274}, 1167 (1996). 

\bibitem{Rosseel}
E.~Rosseel, M.~Van Bael, M. Baert, R.~Jonckheere, V.V.~Moshchalkov, 
and Y.~Bruynseraede, Phys.~Rev.~B {\bf 53},
R2983 (1996).

\bibitem{Jensen1}
C.~Reichhardt, R.T.~Scalettar, G.T.~Zim{\' a}nyi, and N. Gr{\o}nbech-Jensen,
Phys.~Rev.~B {\bf 61}, R11 914 (2000).   

\bibitem{Velez}
M.~V{\' e}lez, D.~Jaque, J.I.~Mart{\' \i}n, F.~Guinea, and J.L.~Vicent, 
Phys.~Rev.~B {\bf 65}, 094509 (2002).

\bibitem{Sur}
R.~Surdeanu, R.J.~Wijngaarden, R.~Griessen, J.~Einfeld, 
and R.~W{\" o}rdenweber, 
Euorphys.~Lett.~ {\bf 54}, 682 (2001).    

\bibitem{Nori}
C. Reichhardt and F.~Nori, Phys.~Rev.~Lett.~{\bf 82}, 414 (1999). 

\bibitem{Velez1}
M.~Velez, D.~Jaque, J.I.~Mart{\' \i}n, M.I.~Montero, I.K.~Schuller, and 
J.L. Vicent, Phys.~Rev.~B {\bf 65}, 104511 (2002).

\bibitem{Sil}
A.V.~Silhanek, L.~Van Look, S.~Raedts, R.~Jonckheere, and V.V.~Moshchalkov,
Phys.~Rev.~B {\bf 68}, 214504 (2003). 

\bibitem{Velez2}
J.E.~Villegas, E.M.~Gonzalez, M.I.~Montero, I.K.~Schuller, and J.L.~Vicent,
Phys.~Rev.~B {\bf 72}, 064507 (2005).   

\bibitem{Wun}
T.C.~Wu, P.C.~Kang, L.~Horng, J.C.~Wu, and T.J.~Yang,
J.~Appl.~Phys. {\bf 95}, 6696 (2004).  

\bibitem{Jiang}
Z.~Jiang, D.A.~Dikin, V.~Chandrasekhar, V.V.~Metlushko, and V.V.~Moshchalkov,
Appl.~Phys.~Lett.~{\bf 84}, 5371 (2004).   

\bibitem{Nishio}
Q.H.~Chen, C.~Carballeira, T.~Nishio, B.Y.~Zhu, and V.V.~Moshchalkov, 
Phys.~Rev.~B {\bf 78}, 172507 (2008). 

\bibitem{Fiory}
A.T.~Fiory, A.F.~Hebard, and S.~Somekh,
Appl.~Phys.~Lett. {\bf 32}, 73 (1978).

\bibitem{Metlushko}
V.V.~Metlushko, M.~Baert, R.~Jonckheere, V.V.~Moshchalkov, and
Y.~Bruynseraede, Solid State Commun. {\bf 91}, 331 (1994); 
M.~Baert, V.V.~Metlushko, R.~Jonckheere, V.V.~Moshchalkov, and
Y.~Bruynseraede, Phys.~Rev.~Lett.~{\bf 74}, 3269 (1995);
M.~Baert, V.V.~Metlushko, R.~Jonckheere, V.V.~Moshchalkov, and
Y.~Bruynseraede, Europhys.~Lett.~{\bf 29}, 157 (1995). 

\bibitem{Welp}
V.~Metlushko, U.~Welp, G.W.~Crabtree, Z.~Zhang, S.R.J.~Brueck,
B.~Watkins, L.E.~DeLong, B.~Ilic, K.~Chung, and P.J.~Hesketh,
Phys.~Rev.~B {\bf 59}, 603 (1999); 
A.A.~Zhukov, P.A.J.~de Groot, V.V.~Metlushko, and B.~Ilic,
Appl.~Phys.~Lett.~{\bf 83}, 4217 (2003);  
U.~Welp, X.L.~Xiao, V.~Novosad, and V.K.~Vlasko-Vlasov,
Phys.~Rev.~B {\bf 71}, 014505 (2005). 

\bibitem{Bending}
A.N.~Grigorenko, S.J.~Bending, M.J.~Van Bael, M.~Lange, V.V.~Moshchalkov,
H.~Fangohr, and P.A.J.~de Groot, Phys.~Rev.~Lett.~{\bf 90}, 237001 (2003).   

\bibitem{Field}
S.B.~Field, S.S.~James, J.~Barentine, V.~Metlushko, G.~Crabtree,
H.~Shtrikman, B.~Ilic, and S.R.J.~Brueck,
Phys.~Rev.~Lett.~{\bf 88}, 067003 (2002).

\bibitem{Commensurate}
C.~Reichhardt, C.J.~Olson, and F.~Nori, Phys.~Rev.~B {\bf 57}, 7937 (1998).  

\bibitem{Peeters1}
G.R.~Berdiyorov, M.V.~Milosevic, and F.M.~Peeters, 
Phys.~Rev.~B {\bf 76}, 134508 (2007). 

\bibitem{Karapetrov}
G.~Karapetrov, J.~Fedor, M.~Iavarone, D.~Rosenmann, and W.K.~Kwok,
Phys.~Rev.~Lett.~{\bf 95}, 167002 (2005);
C.J. Olson Reichhardt, A. Lib{\' a}l, and C. Reichhardt, 
Phys.~Rev.~B {\bf 73}, 184519 (2006). 

\bibitem{Peeters}
G.R.~Berdiyorov, M.V.~Milosevic, and F.M.~Peeters,
Phys.~Rev.~Lett.~{\bf 96}, 207001 (2006); 
G.R.~Berdiyorov, M.V.~Milosevic, and F.M.~Peeters,
Phys.~Rev.~B {\bf 74}, 174512 (2006). 

\bibitem{Pannetier}
A.~Bezryadin, Y.N.~Ovchinnikov, and B.~Pannetier, 
Phys.~Rev.~B {\bf 53}, 8553 (1996).

\bibitem{Jensen}
C.~Reichhardt and N.~Gr{\o}nbech-Jensen, 
Phys.~Rev.~Lett.~{\bf 85}, 2372 (2000). 

\bibitem{Horng}
L.~Horng, T.J.~Yang, R.~Cao, T.C.~Wu, J.C.~Lin, and J.C.~Wu,
J.~Appl.~Phys.~{\bf 103}, 07C706 (2008). 

\bibitem{Wu}
T.C.~Wu, J.C.~Wang, L.~Horng, J.C.~Wu, and T.J.~Yang, 
J. Appl.~Phys.~{\bf 97}, 10B102 (2005). 

\bibitem{Martin}
J.I.~Mart{\' i}n, M.~V{\' e}lez, J.~Nogu{\' e}s, and I.K.~Schuller,
Phys.~Rev.~Lett.~{\bf 79}, 1929 (1997);
A.~Hoffmann, P.~Prieto, and I.K.~Schuller, Phys.~Rev.~B {\bf 61}, 6958 (2000);  
J.I.~Mart{\' \i}n, M.~V{\' e}lez, A.~Hoffmann, I.K.~Schuller,
and J.L.~Vicent,
Phys.~Rev.~Lett.~{\bf 83}, 1022 (1999);
M.J.~Van Bael, J.~Bekaert, K.~Temst, L.~Van Look, V.V.~Moshchalkov,
Y.~Bruynseraede, G.D.~Howells, A.N.~Grigorenko, S.J.~Bending, and G.~Borghs,
Phys.~Rev.~Lett.~{\bf 86}, 155 (2001).

\bibitem{Fertig} 
D.J.~ Priour and H.A.~Fertig, Phys.~Rev.~Lett.~{\bf 93}, 057003 (2004);
Q.H.~Chen, G.~Teniers, B.B.~Jin, and V.V.~Moshchalkov, Phys.~Rev.~B {\bf 73},
014506 (2006). 

\bibitem{Morgan}
D.J.~Morgan and J.B.~Ketterson,
Phys.~Rev.~Lett.~{\bf 80}, 3614 (1998).

\bibitem{Bigelow}
H.~Pu, L.O.~Baksmaty, S.~Yi, and N.P.~Bigelow,
Phys.~Rev.~Lett.~{\bf 94}, 190401 (2005);
J.W.~Reijnders and R.A.~Duine, Phys.~Rev.~A {\bf 71}, 063607 (2005). 

\bibitem{Demler}
A.A.~Burkov and E.~Demler, Phys.~Rev.~Lett.~{\bf 96}, 180406 (2006). 

\bibitem{Tung}
S.~Tung, V.~Schweikhard, and E.A.~Cornell, 
Phys.~Rev.~Lett.~{\bf 97}, 240402 (2006). 

\bibitem{Pinning}
C. Reichhardt and C.J. Olson Reichhardt, 
Phys.~Rev.~B {\bf 76}, 064523 (2007). 

\bibitem{Transverse} 
C. Reichhardt and C.J. Olson Reichhardt, 
Phys.~Rev.~Lett.~{\bf 100}, 167002 (2008).     

\bibitem{Transverse2}
C. Reichhardt and C.J. Olson Reichhardt, 
Phys.~Rev.~B {\bf 78}, 224511 (2008).  

\bibitem{Toner}
L.~Radzihovsky and J.~Toner, Phys.~Rev.~Lett.~{\bf 81}, 3711 (1998). 

\bibitem{Dohmen}
N.~Markovic, M.A.H.~Dohmen, and H.S.J.~van der Zant, 
Phys.~Rev.~Lett.~{\bf 84}, 534 (2000). 

\bibitem{Colloid}
C. Reichhardt and C.J. Olson,  Phys.~Rev.~Lett.~{\bf 88}, 248301 (2002);
M.~Mikulis, C.J.~Olson Reichhardt, C.~Reichhardt, R.T.~Scalettar, 
and G.T.~Zim{\' a}nyi,
J.~Phys: Condens. Matter {\bf 16}, 7909 (2004).   

\bibitem{Brunner} 
M.~Brunner and C.~Bechinger, Phys.~Rev.~Lett.~{\bf 88}, 248302 (2002). 

\bibitem{Trizac}
R.~Agra, F. van Wijland, and E.~Trizac, 
Phys.~Rev.~Lett.~{\bf 93}, 018304 (2004). 

\bibitem{Frey}
A.~Sarlah, E.~Frey, and T.~Franosch, 
Phys.~Rev.~E {\bf 75}, 021402 (2007). 

\bibitem{Ying}
E.~Granato and S.C.~Ying, Phys.~Rev.~B {\bf 69}, 125403 (2004);
J.~Tekic, O.M.~Braun, and B.~Hu, Phys.~Rev.~E {\bf 71}, 026104 (2005); 
L.~Bruschi, G.~Fois, A.~Pontarollo, G.~Mistura, B.~Torre,
F.B.~de Mongeot, C.~Boragno, R.~Buzio, and U.~Valbusa,
Phys.~Rev.~Lett.~{\bf 96}, 216101 (2006).  

\bibitem{Guthermann}
G.~Coupier, M. Saint Jean, 
and C.~Guthmann, Phys.~Rev.~B {\bf 75}, 224103 (2007). 

\bibitem{Grier}
P.T.~Korda, M.B.~Taylor, and D.G.~Grier, 
Phys.~Rev.~Lett.~{\bf 89}, 128301 (2002); 
A.~Gopinathan and D.G.~Grier, 
Phys.~Rev.~Lett.~{\bf 92}, 130602 (2004). 

\bibitem{Dahmen}
J.P.~Sethna, K.A.~Dahmen, 
and C.R. Myers, Nature (London) {\bf 410}, 242 (2001). 

\bibitem{Misko3}
W.V.~Pogosov, A.L.~Rakhmanov, and V.V.~Moshchalkov,
Phys.~Rev.~B {\bf 67}, 014532 (2003). 

\end{thebibliography}
\end{document}